\documentclass[12pt,reqno]{article}
\usepackage{amsmath}
\usepackage{amsthm}
\usepackage{amsfonts}
\usepackage{amssymb}
\usepackage{epsfig}
\usepackage{graphics}
\usepackage{graphicx,indentfirst}
\numberwithin{equation}{section}

\def \Fig#1#2#3 {
\begin{figure}
\centering
\epsfxsize=#2cm \epsfbox{#1.eps}
\caption{#3}
\label{#1}
\end{figure}
}

\newcount\figno
\def\fig#1#2#3{
\par\begingroup\parindent=0pt\leftskip=1cm\rightskip=1cm\parindent=0pt
\baselineskip=15pt
\global\advance\figno by 1
\epsfxsize=#3
\centerline{\epsfbox{#2}}
\vskip 12pt
{\bf \small Figure \the\figno:} {\small #1}\par
\endgroup\par
}
\def\figlabel#1{\xdef#1{\the\figno
\mbox{ }}}
\def\encadremath#1{\vbox{\hrule\hbox{\vrule\kern8pt\vbox{\kern8pt
\hbox{$\displaystyle #1$}\kern8pt}
\kern8pt\vrule}\hrule}}

\def\L2{{\it Fun\/}\bigl(\text{\GL}\bigr)}
\newcommand{\nol}{:\!}
\newcommand{\nor}{\!\!:}
\newcommand{\G}{G}

\newcommand{\Ad}{\text{Ad}}

\def\GL{{\rm GL(1$|$1)}}

\newcommand{\g}{\mathfrak{g}}
\newcommand{\ap}{\mathfrak{a_+}}
\newcommand{\am}{\mathfrak{a_-}}
\newcommand{\apm}{\mathfrak{a_\pm}}
\newcommand{\azero}{\mathfrak{a_0}}

\newcommand{\cN}{\mathcal{N}}

\def\GL{{\rm GL(1$|$1)}}

\newtheorem{thm}{Theorem}[section]

\DeclareMathOperator{\tr}{tr}

\DeclareMathOperator{\str}{str}

\newcommand{\al}{\alpha} \newcommand{\ga}{\gamma}
\newcommand{\Ga}{\Gamma} \newcommand{\be}{\beta}
\newcommand{\ka}{\kappa} \newcommand{\de}{\delta}
\newcommand{\ep}{\epsilon} \newcommand{\si}{\sigma}
 
 \newcommand{\Om}{\Omega}

\newcommand{\half}{\mbox{$\frac{1}{2}$}}

\newcommand{\fourth}{\mbox{$\frac{1}{4}$}}

\newtheorem{example}[thm]{Example}

\newcommand{\ph}{\phantom}

\newcommand{\expect}[1]{\langle #1\rangle}

\newcommand{\normord}[1]{{} : #1 : {}}

\newcommand{\del}{\partial}
\newcommand{\delbar}{\bar{\partial}}

\newcommand{\Id}{\textrm{Id}}

\newcommand{\adsst}{\textrm{AdS}_3\times\textrm{S}^3\times\textrm{T}^4}
\newcommand{\adss}{\textrm{AdS}_3\times\textrm{S}^3}
\newcommand{\ads}{\textrm{AdS}}

\newcommand{\trm}{\textrm}

\begin{document}
\begin{titlepage}

 \renewcommand{\thefootnote}{\fnsymbol{footnote}}
\begin{flushright}
 \begin{tabular}{l}
 DESY 10-098\\
 WITS-CTP-54
 \end{tabular}
\end{flushright}

 \begin{center}

 \vskip 2.5 truecm

\noindent{\large \textbf{From world-sheet supersymmetry to super target spaces}}\\
\vspace{1.5cm}

\noindent{ Thomas Creutzig$^a$\footnote{E-mail: creutzig@physics.unc.edu}
and Peter B. R\o nne$^{b,\, c}$\footnote{E-mail: peter.roenne@wits.ac.za}}
\bigskip

 \vskip .6 truecm
\centerline{\it $^a$Department of Physics and Astronomy, University of North Carolina,}
\centerline{\it Phillips Hall, CB 3255, Chapel Hill, NC 27599-3255, USA}
\bigskip
\centerline{\it $^b$DESY Theory Group, DESY Hamburg} \centerline{\it
Notkestrasse 85, D-22607 Hamburg, Germany}
\bigskip
\centerline{\it $^c$National Institute for Theoretical Physics and Centre
for Theoretical Physics,} \centerline{\it University of the Witwatersrand,
Wits, 2050, South Africa}

 \vskip .4 truecm

 \end{center}

 \vfill
\vskip 0.5 truecm

\begin{abstract}

We investigate the relation between $\mathcal N=(2,2)$
\emph{super}conformal Lie group WZNW models and Lie \emph{super}group WZNW
models. The B-twist of an exactly marginal perturbation of the world-sheet
superconformal sigma model is the supergroup model. Moreover, the
superconformal currents are expressed in terms of  Lie superalgebra
currents in the twisted theory. As applications, we find protected sectors
and boundary actions in the supergroup sigma model. A special example is
the relation between string theory on $\adsst$ in the RNS formalism and
the $\textrm{U}(1,1|2)\times\trm{U}(1|1)\times\trm{U}(1|1)$ supergroup
WZNW model.
\end{abstract}
\vfill
\vskip 0.5 truecm

\setcounter{footnote}{0}
\renewcommand{\thefootnote}{\arabic{footnote}}
\end{titlepage}

\newpage

\tableofcontents

\section{Introduction}

In this note we consider a relation between sigma models on bosonic groups
with $\cN=(2,2)$ world-sheet supersymmetry and models with supergroups as
target spaces via topological twisting.
Our motivation to ask for such a relation and to understand
it in detail comes from two sides.

The first  motivation to study the relation comes from boundary theories
on supergroups. For WZNW models on \mbox{type I} Lie supergroups there
exists a nice prescription to compute correlation functions in the bulk
theory \cite{Schomerus:2005bf,Saleur:2006tf,Gotz:2006qp,Quella:2007hr}.
The WZNW model is equivalent to a model consisting of the WZNW model of
the bosonic subgroup, free fermions and an interaction term that couples
bosons and fermions. The first observation we make, is that the action of
the model without the interaction term resembles the topological twist of
an $\cN=(2,2)$ superconformal field theory.

Now, we would like to have a similar free fermion prescription for the
boundary \mbox{type I} supergroup WZNW model. So far only in the case of
GL(1$|$1) this is known \cite{Creutzig:2008ek}.\footnote{Also for the
OSP(1$|$2) boundary WZNW model such a free fermion realization is known
and used \cite{Creutzig:2010zp}.} There, in addition to the bulk fermions,
one had to introduce an additional fermionic boundary degree of freedom.
Moreover, the boundary screening charge looks like the square root of the
bulk interaction term. These two features are well-known in world-sheet
supersymmetric theories, i.e. in order to preserve $\cN=2$ superconformal
symmetry on the boundary additional boundary fermions plus a factorization
of the bulk superpotential into boundary superpotentials is required
\cite{Warner:1995ay}. We want to understand why we have such a similar
behaviour. Moreover, we would like to use techniques from world-sheet
supersymmetry to find boundary actions and hence a perturbative
description involving free fermions to solve boundary supergroup WZNW
models.

The second hint of the relation came
from non-trivial exact checks of the $\ads_3/\trm{CFT}_2$
correspondence~\cite{Gaberdiel:2007vu,Dabholkar:2007ey}. Here correlation
functions of chiral primary operators in the weak coupling limit of string
theory on $\adsst$ were calculated, and precise agreement was found with
calculations done in the dual two-dimensional conformal field theory. Such
an agreement is at first sight surprising since the computations in the
bulk and on the boundary correspond to different points in moduli space,
and some protection of the correlators must be present.
In~\cite{deBoer:2008ss} the explanation for the boundary side was given.
The argument utilizes that the dual conformal theory has a whole
$\cN=(4,4)$ worth of supersymmetry. Using this extended supersymmetry, the
correlators, which correspond to an $\cN=(2,2)$ chiral ring, can be shown
to be covariantly constant over the total moduli space.

The question is whether we can now explain this from the string theory
side, which only has $\cN=(2,2)$ world-sheet supersymmetry, by finding
some protected sectors. From~\cite{Berkovits:1999im} we know that string
theory on $\adsst$ in the hybrid formalism has a description in terms of
the $\trm{PSU}(1,1|2)$ supergroup sigma model where RR-deformations
correspond to deformations away from the WZNW point. This lead us to the
search for topological sectors in $\trm{PSU}(1,1|2)$. We, however, only
found such sectors in $\trm{U}(1,1|2)$, and in general in $\trm{GL}(N|N)$.
Since a conformal topological sector correspond to the twist of a
world-sheet supersymmetric theory, this suggests a relation between the
$\cN=(2,2)$ world-sheet supersymmetric $\trm{GL}(N)\times\trm{GL}(N)$
sigma model and the $\trm{GL}(N|N)$ supergroup model via twisting. Note
that it is important that the supergroup has superdimension zero since the
world-sheet supersymmetric theory has the same number of bosons and
fermions.

Moreover, for an arbitrary supergroup computing all correlation functions is out of
reach at the moment. One could thus be less ambitious and restrict to
correlators involving fields in a subsector of the theory. A good
subsector is then the cohomology of a BRST-like operator $Q$ such that the
Lagrangian splits into the Lagrangian of a simpler model plus a $Q$-exact
term. The correlators of the cohomology can then be computed in the model
corresponding to the simpler Lagrangian \cite{Candu:2010yg}. In
$\cN=(2,2)$ superconformal models such good subsectors are naturally the
chiral rings. With the relation presented in this note we then get
supergroup analogs of chiral rings, and thus distinguished good sectors of
GL(N$|$N) supergroup models in which we might be able to compute interesting
correlation functions.

The Lie supergroup GL(N$|$N) has various applications in statistical and condensed matter physics,
especially in the context of disordered fermion systems \cite{Guruswamy:1999hi} and the integer Qantum Hall effect \cite{Zirnbauer:1999ua}. The supergroup GL(1$|$1) appeared as a topological twist in the problem of percolation and polymers \cite{Saleur:1991hk}.

What we do in detail is to consider the $\cN=(1,1)$ world-sheet
supersymmetric version of the $\textrm{GL}(N)\times\textrm{GL}(N)$ WZNW
model. Using a particular Manin triple decomposition of the algebra we
explicitly construct an extension to $\cN=(2,2)$ supersymmetry
\cite{Getzler:1993py}. This extended supersymmetry is preserved under
certain deformations of the theory which just add background charges to
certain fields. We then perform a B-twist of the theory to obtain a
conformal theory which is topological when restricting to the BRST
cohomology. By fixing the deformation parameters mentioned before we can
obtain the following: Firstly, the Lagrangian of the twisted theory is the
free fermion resolution \cite{Quella:2007hr} of the $\mathrm{GL}(N|N)$
WZNW model without the boson-fermion interaction term. Secondly, the BRST
current of the topological theory is one of the fermionic affine
supercurrents of the model. Likewise the preimages of the BRST current and
stress-energy tensor in the cohomology can be expressed in terms of the
supercurrents. This in turn means that the full supergroup
$\mathrm{GL}(N|N)$ WZNW model including the boson-fermion interaction term
by twisting is related to a supersymmetric deformation of the $\cN=(2,2)$
world-sheet supersymmetric $\textrm{GL}(N)\times\textrm{GL}(N)$ WZNW
model. We will see that the boson-fermion interaction terms correspond to
an F-term deformation with a chiral field. The principal chiral field deformation is a D-term
type of deformation, but it turns out that it is exact in the BRST-charge.

We also consider the important example of string theory on $\adsst$. We
show that our construction is a novel choice of $\cN=(2,2)$ world-sheet
superalgebra, which only slightly differs from the standard choice in the
supercurrent $G^+$ that becomes the BRST-current after twisting. The other
supercurrent $G^-$ agrees with standard string theory. With the new choice
of superalgebra the string theory is related to the
$\textrm{U}(1,1|2)\times\trm{U}(1|1)\times\trm{U}(1|1)$ supergroup WZNW
model. We also show that all interaction terms in the WZNW model are exact
in the supercharge $G^-$ that fitted with standard string theory. Finally,
for the case with boundary we conjecture a solution to the Warner problem.

The article is organized as follows: In section 2 we will introduce the
models and concepts that we need, this includes an introduction to both
world-sheet supersymmetric models and supergroup WZNW models. Section 3
shows in detail how to get the relation between the world-sheet
supersymmetric $\textrm{GL}(N)\times\textrm{GL}(N)$ WZNW model and the
supergroup $\mathrm{GL}(N|N)$ WZNW model. Here we also consider the
supersymmetric deformations of the $\textrm{GL}(N)\times\textrm{GL}(N)$
WZNW model that corresponds to the boson-fermion interaction terms and the
principal chiral field deformations. String theory on $\adsst$ is considered in section 4
along with the Warner problem for the case with boundary. We conclude with
an outlook in section 5.

\section{Superconformal and supergroup WZNW models}

In this section we introduce the two types of models that we study.
Firstly, we review the construction of $\cN=(2,2)$ superconformal symmetry
in world-sheet supersymmetric WZNW models and, secondly, review Lie
supergroup WZNW models and their free fermion resolution. We also briefly
recall the construction of topological theories by twisting.

\subsection{Supersymmetric WZNW models}

We consider a world-sheet supersymmetric WZNW model whose target space is
a Lie group, $G$. To define the model let us use a superspace notation
where the world-sheet is a $2|2$ dimensional supersurface $\Sigma$
parameterized by complex coordinates $z,\bar z$ and two odd coordinates
$\theta,\bar\theta$. The basic field, $\Ga$, is then a map from this super
world-sheet into the Lie group. It turns out that at the WZNW point, this
field has a nice parametrization~\cite{DiVecchia:1984ep}. Let $\g$ be the
Lie algebra of $G$ and $\{t^a\}$ a basis. Then we define the fermionic Lie
algebra valued fields
\begin{equation}\label{eq:fermionpartners}
 \chi(z,\bar z) \ = \ \chi_a(z,\bar z)t^a\qquad\text{and}\qquad\bar\chi(z,\bar z) \ = \ \bar\chi_a(z,\bar z)t^a\, .
\end{equation}
Exponentiating these fields with an odd parameter gives Lie group valued
fields. Furthermore, let $g$ be a bosonic Lie group valued field then we
can parameterize the superfield $\Ga$ as
\begin{equation}
    \Ga\ = \ \exp (i\theta\chi)\ g \ \exp (-i\bar\theta\bar\chi)\, .
\end{equation}

To define the action of the WZNW model we need to fix some non-degenerate
invariant bilinear form $(\cdot,\cdot)$ on the Lie algebra $\g$ of $G$.
The level $k$ of the model is here absorbed into the definition of the
bilinear form $(\cdot,\cdot)$. The action of the world-sheet
supersymmetric model is given by the standard WZNW action with $g$
replaced by $\Ga$, and integration is over the super world-sheet $\Sigma$.
In our parametrization it can be shown to have the form
\begin{align}\label{eq:N=1action}
        S^{\mathcal N=1}_{\rm WZNW}[\Ga]\ =\  S^{\trm{(ren)}}_{\trm{WZNW}}[g]\ +\ \frac{1}{2\pi}\int d^2z\, (\chi,\bar\del\chi)+
        (\bar\chi,\del\bar\chi) \, .
\end{align}
Here the bosonic part of the WZNW action has been renormalized by the
Killing form $(\cdot,\cdot)_{\trm{Kil.}}$
\begin{align}\label{eq:renormmetricworldsheet}
    (\cdot,\cdot)_{\mathrm{ren}}=(\cdot,\cdot)+\half\, (\cdot,\cdot)_{\trm{Kil.}}
\end{align}
We see that the fermions have been decoupled from the bosonic WZNW model.

The WZNW model~\eqref{eq:N=1action} has an $\mathcal N=(1,1)$
superconformal symmetry by construction. Below we will see that in some
cases this can be enhanced to an $\mathcal N=(2,2)$ superconformal
algebra.

\subsection{A Sugawara-like construction of the superconformal algebra}\label{sec:sugawara}

This construction has been introduced in \cite{Getzler:1993py}.
In this section, we construct the $\mathcal N=2$ superconformal algebra of
the world-sheet super symmetric WZNW models. We restrict to the
holomorphic sector, the anti-holomorphic currents are analogous.

The $\mathcal N=2$ superconformal currents consist of the chiral Virasoro
field $T$, two fermionic fields $G^\pm$ with conformal weight $h_G = 3/2$
and a bosonic U(1) current $U$ with weight one. The $\cN=2$ superconformal
algebra is encoded in the operator product expansions
\begin{equation}
\begin{split}\label{eq:defiN2algebra}
G^+(z)\, G^-(w)&\sim\frac{c/3}{(z-w)^3} +
  \frac{U(w)}{(z-w)^2} + \frac{(T+\frac{1}{2}\del U)(w)}{(z-w)}\, ,
 \\[2mm]
U(z) \, G^\pm(w) & \sim   \frac{\pm G^\pm(w)}{(z-w)}\quad ,\ \
U(z)\, U(w) \, \sim \,  \frac{c/3}{(z-w)^2}\, .
\end{split}
\end{equation}

Given an $\cN=1$ superconformal WZNW model, as described in last
subsection, there is a precise criterium whether it possesses an $\mathcal
N=2$ superconformal symmetry. That is, let $\g$ be the Lie algebra of the
Lie group $G$. Suppose there exist two Lie subalgebras $\apm$ such that
\begin{equation}
 \g = \ \ap\ \oplus\ \am\, .
\end{equation}
Further, assume that $\apm$ are isotropic, i.e. the bi-linear form
vanishes on them, then the world-sheet supersymmetric WZNW model possesses
an $\mathcal N=2$ superconformal symmetry.

To explicitly construct the $\cN=2$ currents we introduce some notation.
Choose a basis $x_i$ of the Lie subalgebra $\ap$. With the help of our
bi-linear form $(.,.)$ we can then fix a dual basis $x^i$ of $\am$ such
that $(x_i,x^j) = \delta_{i}^j$. Our choice of basis implies that the Lie
bracket takes the following form
\begin{equation}
    \begin{split}\label{eq:Maninalgebra}
        [x_i,x_j]\ &= \ {c_{ij}}^kx_k\, ,\\
        [x^i,x^j]\ &= \ {f^{ij}}_kx^k\, ,\\
        [x_i,x^j]\ &= \ {c_{ki}}^jx^k+{f^{jk}}_ix_k\ . \\
    \end{split}
\end{equation}
Here ${c_{ij}}^k$ and ${f^{ij}}_k$ are the structure constants of $\ap$
and $\am$, respectively. The last equation follows from the first two
using the invariance of the bilinear form. For later convenience we define
the element $\rho \in \g$
\begin{equation}\label{eq:rho}
 \rho \ : \, = \ - [x^i,x_i]\ = \ {f^{ik}}_ix_k+{c_{ki}}^ix^k\ = \ \rho^kx_k+\rho_kx^k\, .
\end{equation}

We denote the chiral affine currents corresponding to the generators $x_i$
and $x^i$ by $J_i(z)$ and $J^i(z)$. Using~\eqref{eq:Maninalgebra}
and~\eqref{eq:renormmetricworldsheet} their operator products are
\begin{equation}\label{eq:currentOPE}
    \begin{split}
        J_i(z)J_j(w)\ &\sim \
        \frac{\frac{1}{2}(x_i\,,x_j)_{\trm{Kil.}} }{(z-w)^2}+\frac{{c_{ij}}^kJ_k(w)}{(z-w)}\, ,\\[2mm]
         J_i(z)J^j(w)\ &\sim \
        \frac{{\delta_i}^j+\frac{1}{2}(x_i\,,x^j)_{\trm{Kil.}}}{(z-w)^2}+
        \frac{{f^{jk}}_iJ_k(w)+{{c_{ki}}^j}J^k(w)}{(z-w)}\, ,\\[2mm]
        J^i(z)J^j(w)\ &\sim \
        \frac{\frac{1}{2}(x^i\,,x^j)_{\trm{Kil.}}}{(z-w)^2}+\frac{{f^{ij}}_kJ_k(w)}{(z-w)}\, ,
        \\
\end{split}
\end{equation}
The operator product expansions of the fields $\chi^i$ and $\chi_i$ take
the form
\begin{equation}\label{eq:fermionOPE}
    \begin{split}
        \chi_i(z)\, \chi_j(w)\ &\sim \ 0\, ,\\[2mm]
        \chi_i(z)\, \chi^j(w)\ &\sim \ \frac{{\delta_i}^j}{(z-w)}\, ,\\[2mm]
        \chi^i(z)\, \chi^j(w)\ &\sim \ 0\ .\\
    \end{split}
\end{equation}
All these fermions have conformal weight $h(\chi_i) = h(\chi^i)=1/2$.

We can now write the currents explicitly. The Virasoro tensor $T$ is in
the standard Sugawara form
\begin{equation}
    T(z)\ = \ \frac{1}{2}(\nol J^iJ_i\nor + \nol J_i J^i\nor +
    \nol \del \chi^i\chi_i\nor -\nol \chi^i\del \chi_i\nor )\, ,
\end{equation}
whereas the dimension $3/2$ fermionic currents are
\begin{equation}\label{eq:gplusminus}
    \begin{split}
        G^+(z)\ &= \, J_i\chi^i-\frac{1}{2}{c_{ij}}^k\,\nol
        \chi^i\chi^j\chi_k\nor\, ,\\[2mm]
        G^-(z)\ &= \, J^i\chi_i-\frac{1}{2}{f^{ij}}_k\,
        \nol \chi_i\chi_j\chi^k\nor \, .\\
    \end{split}
\end{equation}
Finally, the dimension one bosonic current $U$ is
\begin{equation}
    U(z)\ = \ \nol \chi^i\chi_i \nor+ \rho^k J_k +
    \rho_k J^k +{c_{mn}}^i{f^{mn}}_j\nol \chi^j\chi_i\nor\, .
\end{equation}
Using the OPEs \eqref{eq:currentOPE} and \eqref{eq:fermionOPE} one can
verify that these currents satisfy the relations of the $\cN=2$
superconformal algebra~\eqref{eq:defiN2algebra}. The anti-chiral partners
$\bar T, \bar G^\pm$ and $\bar U$ are constructed in complete analogy.

\subsection{Deformations}\label{sec:deformations}

The above construction can be slightly generalized~\cite{Getzler:1993py}.
Actually there exist a family of supersymmetric deformations defined by
elements in $\azero$, the orthogonal complement of the direct sum of the
derived subalgebras of $\ap$ and $\am$, i.e.
\begin{equation}
 \azero \ = \ \{x\,\in\,\g\,|\,(x,y)\,=\,0\, \forall\,
 y\,\in\,[\ap,\ap]\,\oplus\,[\am,\am]\}\, .
\end{equation}
Consider an element $\alpha=p^ix_i+q_ix^i \in \azero$. It follows from the
definition of $\azero$ that the components $p^i$ and $q_i$ must satisfy
\begin{equation}\label{eq:alpha}
    {c_{ij}}^kq_k \ = \ {f^{ij}}_kp^k \ = \ 0\, .
\end{equation}
Given the element $\alpha$, the deformed currents of the $\cN=2$
superconformal algebra are as follows
\begin{equation}\label{eq:deformedvirasoro}
    \begin{split}
        U_\alpha(z) \ &= \ U(z)+p^i\, I_i(z)-q_i\, I^i(z)\, ,\\[2mm]
        T_\alpha(z) \ &= \ T(z)+\frac{1}{2}(p^i\, \del I_i(z)+
        q_i\, \del I^i(z))\, .\\
    \end{split}
\end{equation}
Here we used the following set of level $k$ Lie superalgebra currents
\begin{equation*}
    \begin{split}
        I_i \, &= \, J_i-{c_{ij}}^{k} \nol \chi^j\chi_k\nor -\frac{1}{2}{f^{jk}}_i \nol \chi_j\chi_k\nor\, ,\\[2mm]
        I^i \, &= \, J^i-{f^{ij}}_k \nol \chi_j\chi^k\nor -\frac{1}{2}{c_{jk}}^{i}\nol \chi^j\chi^k\nor\, . \\
    \end{split}
\end{equation*}
The expressions for the deformed supercurrents $G^\pm$ are a bit
simpler
\begin{equation}\label{eq:deformedsupercurrents}
    \begin{split}
        G^+_\alpha \ &= \ G^++q_i\, \del \chi^i\, , \\[2mm]
        G^-_\alpha \ &= \ G^-+p^i\, \del \chi_i\, . \\
    \end{split}
\end{equation}
The deformation changes the central charge as
\begin{align}\label{eq:deformedcentralcharge}
c_\alpha = c-6q_ip^i\, .
\end{align}

This deformed $\cN=2$ structure extends a deformation of the original
$\cN=1$ superconformal algebra, and will be important in our discussion.
Note that the deformation simply changes the energy-momentum tensor by
derivatives of the generalized currents $I_i, I^i$. In simple cases this
is just adding background charges to the action. Indeed, as shown in
\cite{Creutzig:2009zz} these deformations are closely related to spectral
flow.

\subsection{Topological conformal field theory}

It will be important for us that an $\cN=2$ superconformal theory
determine topological conformal theories by the twisting procedure. Here
we follow~\cite{Dijkgraaf:1990qw}.

In this paper we only need to consider the positive B-twist. Given an
$\cN=2$ superconformal theory as above, we define the energy-momentum
tensor of the B-twisted theory by
\begin{equation}\label{eq:twistedvirasoro}
\begin{split}
 T^+_{\text{twisted}}(z)\ = \ T(z)+\frac{1}{2}\del U(z)\, ,\\
\bar T^+_{\text{twisted}}(\bar z)\ = \ \bar T(\bar z)+\frac{1}{2}\bar\del \bar U(\bar z)\, .\\
\end{split}
\end{equation}
The twisted theory will by definition have central charge $c=0$. We see,
as in the deformations above, that if we write $U=\del\phi$ this will just
add background charge to $\phi$ in the action, and again it can be seen as
a spectral flow~\cite{Bershadsky:1993cx}. This means that the dimensions
of the fields change and we now have the weights $h_{G^+}=1$, $h_{G^-}=2$
whereas $U$ still have weight one. The twisted theory is not in itself
topological, but if we restrict ourselves to the states in the
BRST-cohomology of $G^+$, we get a conformal topological
theory.\footnote{Had we considered the negative B-twist with
$T^-_{\text{twisted}}(z) = T(z)-\frac{1}{2}\del U(z)$ the BRST operator
would be $G^-$.} Indeed, from the $\cN=2$ algebra~\eqref{eq:defiN2algebra}
we see that the zero modes of $G^++\bar G^+$ satisfy
\begin{equation}
\begin{split}
 (G^+_0+\bar G^+_0)^2 \ &= \ 0\ ,\\
[G^+_0+\bar G^+_0,G^-(z)]\ &= \ T^+_{\text{twisted}}(z)\, ,\\
[G^+_0+\bar G^+_0,\bar G^-(\bar z)]\ &= \ \bar T^+_{\text{twisted}}(\bar z)\, ,\\
\end{split}
\end{equation}
which precisely is the algebra of a conformal topological theory with BRST
charge $Q+\bar Q\equiv G^+_0+\bar G^+_0$. The physical states are defined
by the cohomology
\begin{equation}
 \mathcal{H}_{\text{phys}} \ = \ \frac{\text{kernel}(Q+\bar Q)}{\text{image}(Q+\bar Q)}\, .
\end{equation}
Note that $G^-$ is the preimage of the twisted stress-energy tensor, and
$U$ will be the preimage of the BRST charge $Q$ itself. Using these
relations one can show that the physical correlation functions
\begin{equation}
 \langle \phi_1(z_1,\bar z_1)\dots\phi_n(z_n,\bar z_n)\rangle_\Sigma
\end{equation}
will depend only on the fields $\phi_i$ and the topology of the
world-sheet $\Sigma$, but not on the world-sheet positions $(z_i,\bar
z_i)$. In the topological CFT the operator product expansion of physical
fields takes the particularly simple form
\begin{equation}
 \phi_i\phi_j \ \sim \ {c_{ij}}^k\phi_k\, .
\end{equation}
 \smallskip

\subsection{Supersymmetric Deformations}

It is important to understand the moduli space of deformations preserving the superconformal algebra.

Let us first relate our notation to the notation in \cite{Hori:2003ic}
where the supercharges are denoted (in the Minkowski notation) $Q_\pm$ and
$\bar Q_\pm$ where the index $\pm$ denotes chirality and hence is related
to our bar notation. On the other hand, the bar notation in
\cite{Hori:2003ic} is related to hermitian conjugation of the
supercharges. In our case this corresponds to the $\pm$ superscript. We
have
\begin{align}\label{eq:superalgebra}
    \{G^+_{-1/2},G^-_{-1/2}\}=2L_{-1},\qquad \{\bar G^+_{-1/2},\bar G^-_{-1/2}\}=2\bar L_{-1}
\end{align}
whereas in \cite{Hori:2003ic} the non-zero anti-commutators are (in Minkowski space and with zero central charges for the supersymmetry algebra)
\begin{align}
    \{Q_\pm,\bar Q_\pm\}=H\pm P=-2i\del/\del x^\pm\, .
\end{align}
Taking $z=x^2+i x^1$ and Wick rotating as $x^2=ix^0$ we get $L_{-1}=-\del_z=i \del/\del x^+$ and $\bar L_{-1}=-\del_{\bar z}=i \del/\del x^-$. So we choose the identification
\begin{align}
    Q_+&=iG^-_{-1/2}\, , & \bar Q_+ &=iG^+_{-1/2}\, , \\
    Q_-&=i\bar G^-_{-1/2}\, , & \bar Q_- &=i\bar G^-_{-1/2}\, .
\end{align}

In the superfield formalism we introduce covariant superderivatives
$D_\pm$ and $\bar D_\pm$. A chiral field, $\Phi_{++}$, is a superfield
with $\bar D_{\pm}\Phi_{++}=0$. Correspondingly an anti-chiral field has
$D_\pm \Phi_{--}=0$, a twisted chiral field has $\bar
D_+\Phi_{+-}=D_-\Phi_{+-}=0$, and finally a twisted anti-chiral field has
$D_+\Phi_{-+}=\bar D_-\Phi_{-+}=0$. The component fields of a chiral
superfield form a representation of the supersymmetry algebra.
Specifically we find that a field $\phi_{ab}$ is the lowest component of a
(twisted) (anti-)chiral superfield $\Phi_{ab}$ if and only if
\begin{align}\label{eq:chirallowest}
    [G^a_{-1/2},\phi_{ab}]=[\bar G^b_{-1/2},\phi_{ab}]=0\, .
\end{align}
i.e. if it belongs to the (ab)-chiral ring (i.e. (++) is (cc) etc. in
standard notation). The middle components are then given by
\begin{align}\label{eq:chiralfermions}
    \psi_{ab}=-[G_{-1/2}^{-a},\phi_{ab}]\, ,\quad \bar\psi_{ab}=-[\bar G_{-1/2}^{-b},\phi_{ab}]\, ,
\end{align}
and the highest order F-term component is
\begin{align}\label{eq:F-component}
    F_{ab}=-\{G_{-1/2}^{-a},[\bar G_{-1/2}^{-b},\phi_{ab}]\}\, .
\end{align}
A functional of a chiral field is again a chiral field which is reflected by the ring nature of the chiral ring. We can also take world-sheet derivatives and preserve chirality.

Using these fields we can build actions that, at least classically, are
invariant under the supersymmetry transformations. We have two type of
terms: F-terms which only depend on one type of fields and which always
will have the form~\eqref{eq:F-component}. For F-terms the lowest
component field $\phi_{ab}$ needs to have dimension $(1/2,1/2)$. The
second type of terms are D-terms which consist of different types of
fields and can change the metric or B-field. These terms needs to have
dimension zero.

By~\eqref{eq:superalgebra} we see that under a positive B-twist, the
F-term perturbations generated by anti-chiral, twisted chiral or twisted
anti-chiral superfields are all exact up to total derivatives. That is,
they can be written as $[G^+_{-1/2}+\bar G^+_{-1/2},\phi]+\textrm{total
derivatives}$, for some field $\phi$.

\subsection{Lie supergroup WZNW models}\label{sec:freefermion}

Using the twist procedure from last subsection we want to relate
world-sheet supersymmetric theories to WZNW models of type I Lie
supergroups. In this section we recall results on these from
\cite{Quella:2007hr}.

We consider a type I Lie superalgebra $\g$. Examples of fundamental matrix
representations are as follows:

\begin{example}
$gl(n|m)$ is given by
\begin{equation}
    \begin{split}
        \text{gl}(n|m)=\Bigl\{ \left(\begin{array}{c|c}A & B\\ \hline C & D\\ \end{array}\right) \Bigr\},
    \end{split}
\end{equation}
where the bosonic matrices $A$ and $D$ are square matrices of size
$n\times n$ and $m\times m$, and the odd matrices $B$ and $C$ respectively
are of size $n\times m$ and $m\times n$. The supertrace is a
supersymmetric non-degenerate invariant bilinear form and it is defined
via
\begin{equation}
    \begin{split}
        \str  \left(\begin{array}{c|c}A & B \\ \hline C & D\\ \end{array}\right) =\tr A -\tr D \ .
    \end{split}
\end{equation}
\end{example}
\begin{example}\label{example:slmn}
 $sl(n|m)$
\begin{equation}
    \begin{split}
        \text{sl}(n|m)=\bigl\{ X \ \in \text{gl}(n|m) \ | \ \str X =0\bigr\}\, ,
    \end{split}
\end{equation}
for $n\neq m$. If $n=m$ $sl(n|n)$ is not simple, in this case one obtains
the projective unitary superalgebra $psl(n|n)$ as the quotient of
$sl(n|n)$ by its one dimensional ideal $\mathcal{I}$ generated by the
identity matrix $1_{2n}$, i.e. $psl(n|n)=sl(n|n)/\mathcal{I}$.
\end{example}

Following \cite{Quella:2007hr} we denote the upper fermionic generators, the positive fermionic roots,  by $S_1^a$, the lower fermionic by $S_{2a}$ and the bosonic by $K^i$. As a non-degenerate invariant bilinear form, we use the supertrace
\begin{align}\label{eq:ksupertrace}
    \expect{A,B}\ =\ k\str(AB)\, .
\end{align}
The bosonic part of this metric is denoted $\ka^{ij}$, and the fermionic part is
\begin{align}
    \expect{S_1^a,S_{2b}}=k\de^a_b\, .
\end{align}

We parameterize a supergroup valued field as
\begin{align}\label{eq:gparam}
    g\ =\ e^c g_B e^{\bar c},
\end{align}
where we have introduced fermionic fields $c$ and $\bar c$
\begin{align}\label{eq:ccbar}
    c=c^a S_{2a},\quad \bar c={\bar c}_a S_1^a\,.
\end{align}

The fermions transform in some representation of the bosonic algebra. We introduce the representation matrices $R^i$ by
\begin{align}\label{eq:repmatr}
    [K^i,S_1^a]=-{(R^i)^a}_b S_1^b,
\end{align}
which implies
\begin{align}
    [K^i,S_{2a}]&=S_{2b}{(R^i)^b}_a\, ,\nonumber \\
    [S_1^a,S_{2b}]&=-k{(R^i)^a}_b\ka_{ij}K^j\, .
\end{align}
By $R(g_B)$ we denote the representation of the group element $g_B$.

A first order formalism for the fermions (called the free fermion
resolution~\cite{Quella:2007hr}) is obtained by introducing auxiliary
dimension 1 $b$-ghosts to match the fermionic fields $c$:
\begin{align}
    b=b_a S_1^a,\quad \bar b=\bar b^a S_{2a}.
\end{align}
The WZNW action then becomes
\begin{equation}\label{eq:actionfermionresul}
    \begin{split}
        S^{\textrm{WZNW}}[g] \ &= \ S_0 \ + \ S_{\text{pert}} \\
        S_0 \ &= \ S[g_B]_{\text{ren+dil}} \ + \ \frac{1}{2\pi}\int_\Sigma d\tau d\sigma\ \str(b\bar\del c)-\str(\bar b\del\bar c)\\
        S_{\text{pert}} \ &= \ -\frac{1}{4\pi k}\int_\Sigma d\tau d\sigma\ \str( \Ad(g_B)(\bar b)b)\, ,\\
    \end{split}
\end{equation}
where $S[g_B]_{\text{ren+dil}}$ is a renormalized version of the bosonic
WZNW action plus dilatonic terms. Written out in components we get
\begin{multline}\label{eq:actionfreefermionic}
    S^{\textrm{WZNW}}[g]=S^{\textrm{WZNW}}_{ren}[g_B]-\frac{1}{8\pi} \int dz^2 \sqrt{h}\mathcal{R}^{(2)}\ln\det R(g_B)\\
    +\frac{1}{2\pi}\int dz^2 \left(b_a\delbar c^a-\bar b_a\del \bar c^a-b_a {R(g_B)^a}_b\bar b^b \right),
\end{multline}

The extra terms are due to the change in the quantum measure. This gives
rise to the Fradkin-Tseytlin term where $h$ is the determinant of the
world-sheet metric and $\mathcal R^{(2)}$ is the world-sheet curvature.
Further, there is a renormalization of the metric in the bosonic part
given by
\begin{align}\label{eq:renormmetric}
    \expect{K_B^i,K_B^j}_{ren}=\kappa^{ij}-\ga^{ij},\quad \ga^{ij}=\tr{R^iR^j}.
\end{align}
Here we have denoted the bosonic currents corresponding to this
renormalized metric by $K_B^i$. Even for simple superalgebras $\g$, $\ga$
may not be proportional to $\ka$.

The affine currents now take the following form (factors of $k$ are
absorbed in the metric)
\begin{equation}
 J \ = \ -\del gg^{-1} \ = \ -\del c + K_B + [c,K_B]+\frac{1}{k}b+\frac{1}{k}:[c,b]: +\frac{1}{2k}:[c,:[c,b]:]:\, .
\end{equation}

We have to be careful with signs when writing the currents in components \cite{Creutzig:2009zz}.
In our case of type I superalgebras we have
\begin{equation}
 \begin{split}\label{eq:signsoncurrents}
   J^a(z)\ = \ \left\{\begin{array}{cc}\ (J(z), t^a) &
\text{if}\ t^a\ \text{in}\ \g_B\oplus\g_{+}\\
-(J(z), t^a) & \qquad\text{if}\ t^a\ \text{in}\ \g_{-}\\
\end{array}\right. .
 \end{split}
\end{equation}
This gives us (with $K_B(z)=K_B^i(z)\ka_{ij}K^j$)
\begin{align}
\begin{split}\label{eq:componentscurrents}
    J^{K^i}(z)&=J^{K_B^i}(z)+b_a(z) {(R^i)^a}_b c^b(z), \\
    J^{S_1^a}(z)&=k\del c^a(z)+k{(R^i)^a}_b \ka_{ij}c ^b(z) J^{K_B^j}(z)-\frac{k}{2} {(R^i)^a}_b \ka_{ij} {(R^j)^c}_d b_c c^b c^d(z), \\
    J^{S_{2a}}(z)&=-  b_a(z),
\end{split}
\end{align}
where right-nested normal ordering is understood. These currents satisfy the OPEs
\begin{align}\label{eq:currentopes}
    J^a(z)J^b(w)\sim\frac{k\str(t^at^b)}{(z-w)^2}+\frac{{f^{ab}}_cJ^c}{z-w},
\end{align}
where ${f^{ab}}_c$ are the structure constants. This can be checked using
the OPEs for $K_B^i$ remembering the
renormalization~\eqref{eq:renormmetric}, and the OPEs for the fermions
\begin{align}
    b_a(z)c^a(w)\sim\frac{1}{z-w}.
\end{align}

Finally, the energy-momentum tensor can be written as
\begin{align}
   T^{\textrm{FF}}= \half\left(J^{K_B^i}\Om_{ij}J^{K_B^j}+\tr(\Om R^i)\ka_{ij}\del J^{K_B^j}\right)-b_a\del c^a,
\end{align}
where the full-renormalized metric $\Om$ has the bosonic and fermionic parts:
\begin{align}\label{eq:fullrenormalisedmetric}
    (\Om^{-1})^{ij}&=\ka^{ij}-\ga^{ij}+\half {f^{im}}_n {f^{jn}}_m\, , \nonumber \\
    (\Om^{-1})^a_{\phantom{a}b}&=\de^{a}_b+(R^i\ka_{ij}R^j)^a_{\ph{a}b}\, .
\end{align}
Note, that the bosonic part of the energy-momentum tensor is a deformation
of the Sugawara Virasoro field. This resembles very much the form of the
$\cN=2$ deformations considered in the last section
\eqref{eq:deformedvirasoro} and the twisting~\eqref{eq:twistedvirasoro}.
In the following we will explain the relation.

\section{From world-sheet supersymmetry to supergroups}

In this section we will understand a relation between world-sheet
supersymmetric WZNW models and Lie supergroup WZNW models. We start with a
$\cN=(2,2)$ superconformal GL(N) $\times$ GL(N) WZNW model. We find a
truly marginal operator $\Phi$, i.e. a perturbation that preserves the
superconformal algebra. This operator couples bosonic fields with the
world-sheet fermions. Then we perform a topological B-twist. This twist is
identified with the GL(N$|$N) WZNW model in the form sketched as
\begin{equation}\label{eq:sketch}
 \begin{split}
S_{\text{GL(N)} \times \text{GL(N)}}^{\cN=(2,2)} + \frac{1}{2\pi}\int d\tau d\sigma\ \Phi \ \quad &\substack{\text{B-twist}\\ \longrightarrow} \ \quad S_{\text{GL(N}|\text{N)}} \\
G^+(z) \ \quad &\substack{\text{B-twist}\\ \longrightarrow} \ \quad J^F(z) \\
G^-(z) \ \quad &\substack{\text{B-twist}\\ \longrightarrow} \ \quad \sum J^F(z)J^B(z)+\del J^F(z) \\
U(z) \ \quad &\substack{\text{B-twist}\\ \longrightarrow} \ \quad J^B(z)\, . \\
 \end{split}
\end{equation}
I.e. the twisted action is the supergroup WZNW model action, and the
twisted super currents can be identified with affine Lie super algebra
currents. Here $J^F$ and $J^B$ denote fermionic and bosonic currents in
GL(N$|$N). The goal now is to make the above sketch \eqref{eq:sketch}
precise.

We start by introducing the GL(N$|$N) WZNW model, then we consider the
$\cN=(2,2)$ GL(N) $\times$ GL(N) WZNW model, perform the B-twist and
relate this to the GL(N$|$N) WZNW model. We introduce the boson-fermion
interaction term $\Phi$ and explain its exactly marginality. Finally, we
consider perturbations by the principal chiral field and
show that it is a D-term.

\subsection{Some properties of the GL(N$|$N) WZNW model}

In this section, we extend section~\ref{sec:freefermion} in the special
cases of the Lie supergroups GL(N$|$N). A convenient basis for the Lie
superalgebra gl(N$|$N) is  $\{ E_\ep^{\al\be}, F_\ep^{\al\be}\ | \ 1\leq
\al,\be \leq N,\, \ep=\pm\,\}$, where the generators $E$ are bosonic
 and $F$ are fermionic. Compared to
section~\ref{sec:freefermion} the bosonic indices $i,j,\ldots$ are now
each replaced by a triplet $\binom{\al\be}{\ep}$ and the fermionic indices
$a,b,\ldots$ are each replaced by a doublet $(\al\be)$. The advantage of this notation is
that the invariant bilinear form and the super commutation relations are easy to express.
The metric~\eqref{eq:ksupertrace} takes the form
\begin{align}\label{eq:ksupertraceglnn}
    k\str(E^{\al\be}_{\ep}E^{\al'\be'}_{\ep'})&:=\ka^{\binom{\al\be}{\ep}\binom{\al'\be'}{\ep'}}=k\ep\de_{\ep\ep'}\de^{\al\be'}\de^{\be\al'}\, ,\\
    k\str(F^{\al\be}_{\ep}F^{\al'\be'}_{\ep'})&=k\varepsilon_{\ep\ep'}\de^{\al\be'}\de^{\be\al'},
\end{align}
where $\varepsilon_{\ep\ep'}$ is the antisymmetric symbol with
$\ep_{+-}=1$.

The non-vanishing Lie super algebra relations are
\begin{equation}\label{eq:glnngenerators}
 \begin{split}
  [E_\ep^{\al\be},E_{\ep'}^{\ga\de}] \ &= \ \de_{\ep,\ep'}\left(\delta^{\be\ga}E_\ep^{\al\de}- \delta^{\al\de}E_\ep^{\ga\be}\right)\, , \\
  [E_\ep^{\al\be},F_{\ep'}^{\ga\de}] \ &= \ \de_{\ep,\ep'}\delta^{\be\ga}F_\ep^{\al\de}-\de_{-\ep,\ep'}\delta^{\al\de}F_{\ep'}^{\ga\be}\, , \\
  \{F_\ep^{\al\be},F_{\ep'}^{\ga\de}\} \ &= \ \de_{-\ep,\ep'}\left(\delta^{\be\ga}E_\ep^{\al\de}+\delta^{\al\de}E_{\ep'}^{\ga\be}\right)\, . \\
 \end{split}
\end{equation}
Following~\eqref{eq:ccbar} we define the fermionic fields
\begin{align}\label{eq:bcmatrices}
    c&=c^{\be\al}F_-^{\al\be}, & \bar c&=\bar c^{\al\be}F_+^{\al\be},\\
    b&=b_{\al\be}F_+^{\al\be},& \bar b&=\bar b_{\be\al}F_-^{\al\be}.
\end{align}
These
satisfy the OPEs
\begin{equation}
c^{\al\be}(z)b_{\ga\de}(w) \ \sim \ \frac{\delta_{\al\ga}\delta_{\be\de}}{(z-w)}\, ,
\end{equation}
and correspondingly for the bared fields.

The boson-fermion super commutation relation in
eq.~\eqref{eq:glnngenerators} determines the representation matrices $R$
by~\eqref{eq:repmatr}
\begin{align}\label{eq:glnnreprmatr}
    {\left(R^{\binom{\al\be}{\ep}}\right)^{\ga\de}}_{\ga'\de'}=-\de_{\ep,+}\de^{\be\ga}\de^\al_{\ga'}\de^{\de}_{\de'}+\de_{\ep,-}\de^{\al\de}\de^\ga_{\ga'}\de^{\be}_{\de'}\, .
\end{align}
This gives the correction to the bosonic metric from decoupling the
fermions~\eqref{eq:renormmetric}
\begin{align}\label{eq:glnngamma}
    \ga^{\binom{\al\be}{\ep}\binom{\ga\de}{\ep'}}=\de_{\ep,\ep'}N\de^{\al\de}\de^{\be\ga}-\de_{-\ep,\ep'}\de^{\al\be}\de^{\ga\de}\, ,
\end{align}
and the bosonic Killing metric ${f^{im}}_n {f^{jn}}_m$ is
\begin{align}\label{eq:glnnboskilling}
    \ka^{\binom{\al\be}{\ep}\binom{\ga\de}{\ep'}}_{\textrm{Killing}}=2\de_{\ep,\ep'}\left(N\de^{\al\de}\de^{\be\ga}-\de^{\al\be}\de^{\ga\de}\right)\, .
\end{align}

The GL(N$|$N) currents~\eqref{eq:componentscurrents} are then given by
\begin{align}\label{eq:glnncurrents}
    \begin{split}
      J^{E^{\al\be}_{\ep}} &= J^{E_{\ep}^{\al\be}}_B-\de_{\ep,+} b_{\be\ga}c^{\al\ga}+\de_{\ep,-}b_{\ga\al}c^{\ga\be}\, , \\
       J^{F_-^{\al\be}} &=-b_{\be\al} \, ,\\
       J^{F_{+}^{\al\be}} &=k\del c^{\al\be}-c^{\ga\be}J^{E_{+}^{\al\ga}}_B-c^{\al\ga}J^{E_{-}^{\ga\be}}_B-b_{\ga\de}c^{\ga\be}c^{\al\de}.
    \end{split}
\end{align}
$J_B$ denote the bosonic currents with the renormalized
metric~\eqref{eq:renormmetric}.

Similarly the anti-holomorphic currents are
\begin{align}\label{eq:antichiralglnncurrents}
    \begin{split}
      {\bar J}^{E^{\al\be}_{\ep}} &= \bar J^{E_{\ep}^{\al\be}}_B+\de_{\ep,+} \bar b_{\al\ga}\bar c^{\be\ga}-\de_{\ep,-}b_{\ga\be}\bar c^{\ga\al}\, , \\
       \bar J^{F_-^{\al\be}} &=k\del \bar c^{\be\al}+\bar c^{\ga\al}\bar J^{E_{+}^{\ga\be}}_B+\bar c^{\be\ga}\bar J^{E_{-}^{\al\ga}}_B-\bar b_{\ga\de}\bar c^{\ga\al}\bar c^{\be\de}, \\
       \bar J^{F_{+}^{\al\be}} &=\bar b_{\al\be}\,.
    \end{split}
\end{align}

We also need the energy-momentum tensor. The bosonic and
fermionic fully-renormalized metrics~\eqref{eq:fullrenormalisedmetric} are
given by
\begin{align}\label{eq:glnnfullrenormalisedmetric}
    (\Om^{-1})^{\binom{\al\be}{\ep}\binom{\ga\de}{\ep'}}&=k\ep\de_{\ep\ep'}\de^{\al\de}\de^{\be\ga}-\ep\ep'\de^{\al\be}\de^{\ga\de}\, , \nonumber \\
    (\Om^{-1})^{(\al\be)}_{\phantom{\al\be}(\ga\de)}&=\de^{\al}_\ga\de^{\be}_{\de}\, .
\end{align}
The holomorphic component of the stress tensor is then
\begin{align}\label{eq:glnnstressenergy}
    T^{\textrm{FF}}=\half J^{E^{\al\be}_{\ep}}_B\Om_{\binom{\al\be}{\ep}\binom{\ga\de}{\ep'}}J^{E^{\ga\de}_{\ep'}}_B-\frac{N}{2k}
\sum_\al\left(\del J^{E_+^{\al\al}}_B+\del J^{E_-^{\al\al}}_B\right)-b_{\al\be}\del c^{\al\be}.
\end{align}

\subsection{The $\cN=(2,2)$ GL(N) $\times$ GL(N) WZNW model}

We now consider the world-sheet supersymmetric GL(N) $\times$ GL(N) WZNW
model that is related to the GL(N$|$N) WZNW model.

We denote the generators of $\g =$ gl(N) $\oplus$ gl(N) by $E_{\si}^{\al\be}$, $\si=\pm$, with relations
\begin{equation}\label{eq:glnglngenerators}
 \begin{split}
  [E_\ep^{\al\be},E_{\ep'}^{\ga\de}] \ &= \ \de_{\ep,\ep'}\left(\delta^{\be\ga}E_\ep^{\al\de}- \delta^{\al\de}E_\ep^{\ga\be}\right)\, . \\
 \end{split}
\end{equation}
Furthermore we start with a metric given by
\begin{equation}\label{eq:glnncailbmetric}
\ka_{\textrm{start}}(E^{\al\be}_{\ep},E^{\ga\de}_{\ep'}) \ = \
    \ka_{\textrm{start}}^{\binom{\al\be}{\ep}\binom{\ga\de}{\ep'}}\ =\
k\ep\de_{\ep\ep'}\de^{\al\de}\de^{\be\ga}-\ep\ep'\de^{\al\be}\de^{\ga\de}\, . \\
\end{equation}
This is the same as the fully-renormalized metric in
equation~\eqref{eq:glnnfullrenormalisedmetric}. Note that this metric only
differs from the standard trace metric in the $\trm{U}(1)$ parts. Thus it
really only implies a simple field redefinition in the $\trm{U}(1)$
fields. The metric is chosen so that after decoupling the fermions, we get
the metric of the free fermion resolution~\eqref{eq:renormmetric}.

Define the Manin triple $(\g, \ap, \am)$ corresponding
to our starting metric as
\begin{equation}\label{eq:maninglngln}
 \begin{split}
  \g \ &= \ \ap \oplus \am\\
  \ap \ &= \ \text{span}\bigl\{\ x_{\al\be}\ | \ x_{\al\be}\,=\,E_+^{\al\be}+E_-^{\al\be}\ \bigr\}\\
  \am \ &= \ \text{span}\bigl\{\ x^{\al\be} \  | \ x^{\al\al}\,=\,\frac{E_+^{\al\al}-E_-^{\al\al}}{2k}+\frac{1}{2k^2}\Id,
\ x^{\be\al}\,=\,\frac{E_+^{\al\be}}{k}\\
&\qquad\qquad\qquad\qquad\qquad\qquad\qquad\text{and} \ x^{\al\be}\,=\,-\frac{E_-^{\be\al}}{k}\ \text{for} \ \al>\be\ \bigr\}\, .\\
 \end{split}
\end{equation}
where $\Id=\Id_++\Id_-$ denotes the central element given by the sum of the two
u(1) generators $\Id_{\ep}=\sum_{\al} E^{\al\al}_\ep$. Note that $\ap$ forms the Lie algebra
gl(N) and $\am$ is a solvable Lie subalgebra of $\g$ and both
are isotropic. Hence we have a Manin triple. As a basis for $\ap$ and $\am$ we use the $x_{\al\be}$ and $x^{\al\be}$
introduced in \eqref{eq:maninglngln}.
Recall that constructions of the $\mathcal N=2$ superconformal algebra are parameterized by
the orthogonal complement of the direct sum of the derived subalgebras of $\apm$. This is
\begin{align}\label{eq:glnnazero}
    \azero=\text{span}\{x_{\al\al},\sum_\al x^{\al\al}\}\, .
\end{align}
For us the choice of $\cN=2$ superconformal algebra given by $\gamma$ in
$\azero$
\begin{equation}\label{eq:defpar}
 \gamma\ = \ -k\sum_\al x^{\al\al}-\sum_\alpha\frac{1}{2k}\left(2N-2\al+1\right)x_{\al\al}
\end{equation}
is important. It implies \eqref{eq:deformedcentralcharge} that the central
charge is $c_\gamma=0$. In order to display the superconformal structure
following sections \ref{sec:sugawara} and \ref{sec:deformations} we need
two more ingredients. First, the structure constants of $\ap\cong\ $gl(N)
are denoted ${c_{ab}}^c$. Those of $\am$ we call ${f^{ab}}_c$ and they can
be extracted from \eqref{eq:glnglngenerators}. Second the element $\rho$
(see ~\eqref{eq:rho}) is
\begin{align}\label{eq:glnnrhotilde}
    \rho=\frac{1}{k}\sum_{\al}(N-2\al+1)x_{\al\al}.
\end{align}

Further, introduce the bosonic currents $J_{\al\be}$ of $\ap$,
$J^{\al\be}$ of $\am$, the and fermionic fields $\chi_{\al\be}$, and their
partners $\chi^{\al\be}$ with OPEs
\begin{equation}
\chi_{\al\be}(z)\chi^{\ga\de}(w) \ \sim \ \frac{\delta_{\al}^\ga\delta_{\be}^\de
}{(z-w)}\, .
\end{equation}
The $\cN=2$ superconformal algebra can now be written down using sections
\ref{sec:sugawara} and \ref{sec:deformations}, we will display them in a
moment after a performing the B-twist. The B-twist changes the Virasoro
field by a linear dilaton term \eqref{eq:twistedvirasoro}. This implies
that in the twisted theory the fermions $\chi^{\al\be}$ have conformal
dimension zero while the $\chi_{\al\be}$ have weight one. Thus we can
identify the fermions $\chi$ with the $bc$ ghosts of the free fermion
resolution. We choose to identify as
\begin{align}\label{eq:glnnghostcomparison}
    c^{\al\be}=-\chi^{\be\al},\quad b_{\al\be}=-\chi_{\be\al}.
\end{align}
Moreover, we identify the bosonic currents with the bosonic currents from
the free fermion resolution of GL(N$|$N) which have the same OPEs due to
our choice of metric \eqref{eq:glnncailbmetric}. The two stress-energy
tensors then match due to our choice of the deformation parameter $\gamma$
in \eqref{eq:defpar}. This means that the action of the twisted
topological theory is the same as the free fermion resolution action
without the boson-fermion interaction term, i.e. the action $S_0$ in
\eqref{eq:actionfermionresul}.

We can now show that the twisted world-sheet supersymmetric currents are
expressed by affine superalgebra currents. First consider the deformed
supersymmetry current $G^+_\gamma$ (which after twisting has conformal
dimension one) given by \eqref{eq:deformedsupercurrents}
\begin{align}\label{eq:glnngplus}
    G^+_\gamma=J_{\al\be} \chi^{\al\be}-\half {c_{\al\be,\delta\rho}}^{\sigma\lambda} \chi^{\al\be} \chi^{\delta\rho} \chi_{\sigma\lambda} -k\sum_{\al}\del \chi^{\al\al}.
\end{align}
Comparing with~\eqref{eq:glnncurrents} we observe that $G^+_\gamma$ is a
fermionic gl(N$|$N) superalgebra current\footnote{To get the comparison of
the stress-energy tensor of the $\trm{GL}(N)\times\trm{GL}(N)$ and
GL$(N|N)$ models we had to fix all the parameters in
$\ga$~\eqref{eq:defpar} except one. Getting this relation fixes the last
parameter.}
\begin{align}\label{eq:gpluscomparison}
    G^+_\gamma=\sum_{\al}J^{F^{\al\al}_{+}}.
\end{align}
The U(1) current is by~\eqref{eq:deformedvirasoro} given by
\begin{align}\label{eq:glnnu1}
    U_\gamma=\frac{1}{2k}\sum_{\al}(-2\al+1)J_{\al\al}+k\sum_\al J^{\al\al}+\chi^{\al\be}\chi_{\al\be}+\frac{1}{k}\sum_{\al,\be}(\al-\be)\chi^{\al\be}\chi_{\al\be}\, .
\end{align}
Again comparison with~\eqref{eq:glnncurrents} gives the following
identification with a bosonic gl$(N|N)$ superalgebra current
\begin{align}\label{eq:u1comparison}
    U_\gamma=\frac{1}{2k}\sum_{\al}(k+N+1-2\al)J^{E_{+}^{\al\al}}+\frac{1}{2k}\sum_{\al}(-k+N+1-2\al)J^{E_{-}^{\al\al}}\, .
\end{align}
Finally, the $G^-$ current of conformal weight two which is given
by~\eqref{eq:deformedsupercurrents} can be written as
\begin{equation}\label{eq:gminuscomparison}
 \begin{split}
    G_\gamma^-\ =\ &\frac{1}{2k}\sum_{\al}J^{F_{-}^{\al\al}}\Big(J^{E^{\al\al}_{+}}-J^{E^{\al\al}_-}+\frac{1}{k}\sum_{\be}
(J^{E^{\be\be}_+}+J^{E^{\be\be}_-})\Big)-\frac{1}{k}\sum_{\al>\be}J^{{F}_{-}^{\al\be}}J^{E^{\be\al}_-}\\
&+\frac{1}{k}\sum_{\al<\be}J^{F_-^{\al\be}}J^{E^{\be\al}_+}-\frac{1}{2k}\sum_\al(2N-2\al+1)\del J^{F_-^{\al\al}}\, .\\
\end{split}
\end{equation}
Note that this expression is normal ordered which is important for the
coefficients of the derivative terms.

So far we only considered the holomorphic currents. Let us now state the analogous results for the right-moving part.
We choose to use the same Manin decomposition. The deformation parameter $\bar\gamma$ is slightly different,
\begin{align}\label{eq:barglnndeformation}
   \bar \gamma=   \sum_\al \bigl(-\frac{1}{2k}\left(-2\al+1\right)x_{\al\al}+k x^{\al\al}\bigr)\,.
\end{align}
We now perform again the +-twist on the right-moving currents and identify the superconformal currents
with affine superalgebra currents.
First, the energy-momentum tensors match.
The identification of the ghost system in this case is
\begin{align}\label{eq:glnnbarghostcomparison}
    \bar c^{\al\be}=\bar \chi^{\al\be},\quad \bar b_{\al\be}=\bar \chi_{\al\be},
\end{align}
to get a matching between the supersymmetric currents and the super affine currents.

Identification of the currents are now
\begin{align}\label{eq:bargpluscomparison}
    \bar G^+_{\bar \gamma}=\sum_{\al}\bar J^{F^{\al\al}_{-}},
\end{align}
\begin{align}\label{eq:baru1comparison}
    \bar U_{\bar \gamma}=\frac{1}{2k}\sum_{\al}(-k+N+1-2\al)\bar J^{E_{+}^{\al\al}}+\frac{1}{2k}\sum_{\al}(k+N+1-2\al)
\bar J^{E_{-}^{\al\al}},
\end{align}
\begin{equation}\label{eq:bargminuscomparison}
 \begin{split}
    \bar G^-_{\bar \gamma}\ =\ &\frac{1}{2k}\sum_{\al}\bar J^{F_{+}^{\al\al}}
\Big(\bar J^{E^{\al\al}_{+}}-\bar J^{E^{\al\al}_-}+\frac{1}{k}\sum_{\be}(\bar J^{E^{\be\be}_+}+
\bar J^{E^{\be\be}_-})\Big)-\frac{1}{k}\sum_{\al>\be}\bar J^{F_{+}^{\al\be}}\bar J^{E^{\be\al}_-}\\
&+\frac{1}{k}\sum_{\al<\be}\bar J^{F_+^{\al\be}}\bar J^{E^{\be\al}_+}-\frac{1}{2k}\sum_\al(-2\al+1)\del \bar J^{F_+^{\al\al}}\, .\\
\end{split}
\end{equation}

Let us summarize this section. We started with a special choice of
$\cN=(2,2)$ superconformal algebra for the GL(N) $\times$ GL(N) super WZNW
model parameterized by deformation parameters $(\gamma,\bar\gamma)$ in
\eqref{eq:defpar} and \eqref{eq:barglnndeformation}. We then have shown
that the B-twisted topological field theory of this model is embedded in
the free fermion resolution of the GL(N$|$N) WZNW model. Moreover the
BRST-current as well as the ghost-number U(1)-current become affine Lie
superalgebra currents. Also the ghost partner of the BRST-current has a
nice expression in terms of Lie superalgebra currents. The full supergroup
WZNW model is realized by coupling the bosons and fermions
\eqref{eq:actionfermionresul}. Thus, the next step is to understand this
boson-fermion interaction term. Since the world-sheet supercurrents are
expressed in terms of the Lie superalgebra currents, it is a
supersymmetric deformation and indeed it will turn out to be an F-term.

\subsection{The boson-fermion interaction term}\label{sec:bosonfermion}

The boson-fermion interaction term~\eqref{eq:actionfreefermionic} is, as
mentioned above, a supersymmetric deformation of our $\cN=2$
$\mathrm{GL}(N)\times\mathrm{GL}(N)$ WZNW model. Since it is a potential
term, we expect it to be an F-term and since we do not expect it to be
exact, it should be a chiral F-term. On the $\mathrm{GL}(N|N)$ side the
term takes the form
\begin{align}\label{eq:bosonfermioninteraction}
    S_{\text{pert}} \ &= \ -\frac{1}{4\pi k}\int_\Sigma d\tau d\sigma\ \str( \Ad(g_B)(\bar b)b)=\frac{1}{4\pi k}\int_\Sigma d\tau d\sigma\ \bar b_{\be\al}b_{\si\de}\tr(E^{\al\be}A^{-1}E^{\si\de}B)
\end{align}
where we have used the form of the $b$-matrices~\eqref{eq:bcmatrices}
\begin{align*}
    b&=b_{\al\be}F_+^{\al\be},& \bar b&=\bar b_{\be\al}F_-^{\al\be},
\end{align*}
and written
\begin{align}
    g_B=\begin{pmatrix}
          A & 0 \\
          0 & B \\
        \end{pmatrix}.
\end{align}
The $E^{\al\be}$ denotes the basis~\eqref{eq:glnglngenerators} of gl(N).

This boson-fermion interaction term is the F-term of a chiral field corresponding to chiral operator
\begin{align}\label{eq:chiralfield}
    \phi=\tr(A^{-1}B).
\end{align}
Indeed this field satisfies~\eqref{eq:chirallowest},
\begin{align}\label{}
    G^+_\gamma(z)\tr(A^{-1}B)(w)\sim 0,\qquad \bar G^+_{\bar\gamma}(z)\tr(A^{-1}B)(w)\sim 0
\end{align}
using that in the bosonic subgroup $g(z,\bar z)$ transforms as
\begin{align}\label{eq:currentong}
    J_a (z) g(w)\sim \frac{- t_a g}{z-w},\qquad \bar J_a (z) g(w)\sim  \frac{g t_a}{z-w},
\end{align}
and the form of the generators in $\ap$~\eqref{eq:maninglngln}.
Further, it can be shown to be a Virasoro primary of dimension $(1/2,1/2)$.

Using~\eqref{eq:chiralfermions} and the basis of
$\am$~\eqref{eq:maninglngln} we get the fermionic part of the chiral
superfield
\begin{align}\label{}
    \psi=-\frac{1}{k}\chi_{\de\si}\tr\left(A^{-1}E^{\si\de}B\right),\quad\bar \psi=\frac{1}{k}\bar \chi_{\al\be}\tr\left(E^{\be\al}A^{-1}B\right).
\end{align}
The F-term~\eqref{eq:F-component} takes the form
\begin{align}\label{}
    F=-\frac{1}{k^2}\bar \chi_{\al\be}\chi_{\de\si}\tr\left(E^{\be\al}A^{-1}E^{\si\de}B\right).
\end{align}
Using the identification of the ghosts~\eqref{eq:glnnghostcomparison} and~\eqref{eq:glnnbarghostcomparison} we see that the F-term is proportional to the boson-fermion interaction term~\eqref{eq:bosonfermioninteraction} and using the chiral superfield $\Phi$ based on $\phi$ in~\eqref{eq:chiralfield} we have
\begin{align}\label{eq:bosfermchiralfield}
    S_{\text{pert}} = \frac{k}{4\pi}\int_\Sigma d\tau d\sigma\, \Phi\, .
\end{align}

Finally we would like to remark, that the construction gives us a wide range of chiral operators.
One just needs to find a chiral field $\tilde\phi$  of conformal dimension $(1/2,1/2)$
satisfying
\begin{equation}
 G^+_\gamma(z)\tilde\phi(w,\bar w)\ \sim\ \bar G^+_{\bar\gamma}(\bar z)\tilde\phi(w,\bar w)\ \sim\ 0\, .
\end{equation}

\subsection{The principal chiral field as a D-term}

Deformations that change the coefficient of the principal chiral field play an important role in the $\trm{PSU}(1,1|2)$ supergroup sigma model where they describe Ramond-Ramond perturbations of the string theory
\cite{Bershadsky:1999hk,Quella:2007sg}. In our GL$(N|N)$ case the principal chiral field field is given by
\begin{equation}
 S_{\text{kin}}\ = \ -\frac{k}{4\pi}\int\,d^2z\,\expect{g^{-1}\del g,g^{-1}\delbar g}\equiv\int\,d^2z\,\Phi_{\trm{principal}}\, .
\end{equation}
Since this is a kinetic term, we expect it to be a D-term. Indeed, using
the supergroup version of \eqref{eq:currentong}, \eqref{eq:currentopes}
and
 $J=-k\del g g^{-1}$, $\bar J=kg^{-1}\delbar g$ one can compute that
\begin{align}\label{}
    \phi=-\frac{1}{4\pi k}\expect{    \left(\begin{array}{c|c}
    0 & \Id \\
    \hline
    \Id  & 0
    \end{array}\right)g^{-1}\left(\begin{array}{c|c}
    0 & \Id \\
    \hline
    \Id  & 0
    \end{array}\right)J g\bar J   },
\end{align}
is the preimage of the principal chiral field:
\begin{align}\label{}
    \Phi_{\trm{principal}}=-\{G_{-1/2}^{+},[\bar
G_{-1/2}^{+},\phi]\}\, .
\end{align}
$\phi$ is not $G^-$ closed and this is not an F-term, but rather a D-term.
It is, however, $G^++\bar G^+$ exact.

\section{Examples and Applications}

In this section, we give some selected examples and applications.

\subsection{GL(2$|$2) and screening charges as chiral perturbations}

We start by considering the example of the GL(2$|$2) WZNW model from GL(2)
$\times$ GL(2). In supergroup WZNW models one goes, in practice, beyond
the free fermion realization and also introduces a Wakimoto free field
realization for the bosonic subgroup. This is then supplemented with
bosonic screening charges. In this section we will find that in addition
to the boson-fermion interaction term also the bosonic screening charge is
an F-term. As a consequence, the cohomology of $G^-$ can be computed in
free field theory. We start with some explicit formulae.

We denote the generators of gl(2) $\oplus$ gl(2) by $E_\pm^{\al\be}$ as before.
Then the two sl(2)s are generated by
\begin{equation}\label{eq:newbasisgl22}
 K^z_\pm \ = \ E_\pm^{11}-E_\pm^{22} \quad, \quad K^+_\pm \ = \ E_\pm^{12} \quad\text{and} \quad K^-_\pm \ = \ E_\pm^{21}\,
\end{equation}
and the two central elements are
\begin{equation}
 K^0_\pm \ = \ E_\pm^{11}+E_\pm^{22}\, .
\end{equation}

\subsubsection*{The standard Manin decomposition}

We are going to display the Manin decomposition and then the resulting supercurrents in the topological twisted
theory as Lie superalgebra currents.

The bases for the Manin decomposition are
\begin{align}\label{eq:manin}
    \ap&=\Bigl\{\frac{1}{2}(K_+^0+K_+^z+K_-^0+K_-^z),\,\frac{1}{2}(K_+^0-K_+^z+K_-^0-K_-^z),\nonumber\\
&\qquad\qquad\qquad\qquad\qquad\qquad\qquad\qquad\qquad\,K_+^++K_-^+,\,K_+^-+K_-^- \Bigr\} \nonumber\\
    \am&=\Bigl\{\frac{1}{4k}(K_+^0+K_+^z-K_-^0-K_-^z)+\frac{1}{2k^2}\left(K_+^0+K_-^0\right),\,\\ \nonumber
    &\qquad\frac{K_+^0-K_+^z-K_-^0+K_-^z}{4k}+\frac{1}{2k^2}\left(K_+^0+K_-^0\right),\,\frac{1}{k}K_+^-,\,-\frac{1}{k}K_-^+\Bigr\}\, .
\end{align}
The deformation parameter is
\begin{align}
    \gamma=-\frac{1}{2k}\left(K_+^z+K_-^z+(k+4)K_+^0+(4-k)K_-^0\right).
\end{align}

The supercurrents in the twisted theory in terms of superalgebra currents are
\begin{align}
    G^+_\gamma&=J^{F_+^{11}}+J^{F_+^{22}},\\
    G^-_\gamma&=\frac{1}{4k}J^{F_-^{11}}\left(J^{K_+^z}+J^{K_+^0}-J^{K_-^z}-J^{K_-^0}+\frac{2}{k}(J^{K_+^0}+J^{K_-^0})\right)\nonumber\\
    &\qquad +\frac{1}{4k}J^{F_-^{22}}\left(J^{K_+^z}-J^{K_+^0}-J^{K_-^z}+J^{K_-^0}+\frac{2}{k}(J^{K_+^0}+J^{K_-^0})\right)  \nonumber\\
    &\qquad -\frac{1}{k}J^{F_-^{21}}J^{K_-^+}+\frac{1}{k}J^{F_-^{12}}J^{K_+^-}-\frac{3}{2k}\del J^{F_-^{11}}-\frac{1}{2k}\del J^{F_-^{22}},\\
    U_\gamma&=\frac{1}{2}J^{K_+^0}-\frac{1}{2}J^{K_-^0}+\frac{1}{2k}J^{K_+^z}+\frac{1}{2k}J^{K_-^z}.
\end{align}

\subsubsection*{Another Manin decomposition}

Depending on the real form one wants to consider there are particularly
good choices of $\cN=2$ superconformal structure. We will now display a
Manin triple that is well suited for $\adsst$. Again the relation to the
supergroup WZNW model works nicely. The Manin decomposition is given by
\begin{align}\label{eq:gl22manin2}
    \ap&=\{K_+^0+K_-^0,\,K_+^z-K_-^z,\,-K_+^++K_-^-,\,-K_+^-+K_-^+ \} \nonumber\\
    \am&=\{\frac{2+k}{4 k^2}K_+^0-\frac{-2+k}{4 k^2}K_-^0,\,\frac{1}{4 k}(K_+^z+K_-^z),\,-\frac{1}{k}K_-^+,\,-\frac{1}{k}K_+^+\}
\end{align}
The deformation parameter $\gamma$ is in this case
\begin{align}\label{eq:aldeformmanin2}
    \gamma=-\frac{1}{2k}\left(-K_+^z+K_-^z+(k+4)K_+^0+(-k+4)K_-^0\right).
\end{align}
This changes the central charge from 12 to zero.

Further, the fermions in the twisted theory are identified with the
$bc$-ghosts of the free fermion realization of the supergroup as follows,
\begin{align}
    b_{11}&=\chi_4, & b_{12}&=\half(\chi_1+\chi_2), & b_{21}&=\half(-\chi_1+\chi_2), & b_{22}&=-\chi_3, \\
    c^{11}&=\chi^4, & c^{12}&=\chi^1+\chi^2, & c^{21}&=-\chi^1+\chi^2, & c^{22}&=-\chi^3.
\end{align}
Here $\chi_i$ ($\chi^i$) is the fermion corresponding to the $i$-th
generator of $\ap$ ($\am$), see~\eqref{eq:fermionOPE}. The supercurrents
are identified as
\begin{align}\label{eq:gminusnontypic}
    G^+_\gamma&=-J^{F_+^{12}}+J^{F_+^{21}},\\
    G^-_\gamma&=-\frac{1}{4k}J^{F_-^{21}}\left(J^{K_+^z}+J^{K_+^0}+J^{K_-^z}-J^{K_-^0}+\frac{2}{k}(J^{K_+^0}+J^{K_-^0})\right)\nonumber\\
    &\qquad +\frac{1}{4k}J^{F_-^{12}}\left(-J^{K_+^z}+J^{K_+^0}-J^{K_-^z}-J^{K_-^0}+\frac{2}{k}(J^{K_+^0}+J^{K_-^0})\right)  \\
\nonumber    &\qquad -\frac{1}{k}J^{F_-^{22}}J^{K_-^+}+\frac{1}{k}J^{F_-^{11}}J^{K_+^+}-\frac{3}{2k}\del J^{F_-^{12}}
+\frac{1}{2k}\del J^{F_-^{21}},\\
    U_\gamma&=\frac{1}{2}J^{K_+^0}-\frac{1}{2}J^{K_-^0}-\frac{1}{2k}J^{K_+^z}+\frac{1}{2k}J^{K_-^z}.
\end{align}\
The anti-holomorphic side is analogous if we choose the Manin triple for
the anti-holomorphic side to be different, but only by an automorphism. It
is
\begin{align}\label{eq:gl22manin2antiholo}
    \ap&=\{-K_+^0-K_-^0,\,-K_+^z+K_-^z,\,K_+^--K_-^+,\,K_+^+-K_-^- \} \nonumber\\
    \am&=\{-\frac{2+k}{4 k^2}K_+^0+\frac{-2+k}{4 k^2}K_-^0,\,-\frac{1}{4 k}(K_+^z+K_-^z),\,\frac{1}{k}K_-^-,\,\frac{1}{k}K_+^-\}
\end{align}
The deformation parameter coincides with the one of the chiral half, $\bar\gamma=\gamma$.

\subsubsection*{Screening charges as chiral perturbations}

We will now show that the screening charges -- both the boson-fermion
interaction terms and the bosonic screening charges can be seen as chiral
perturbations.

For the standard Manin decomposition we have already seen that the
boson-fermion interaction terms can be seen as a chiral
perturbation~\eqref{eq:bosfermchiralfield}. This also happens in the case
of the alternative Manin decomposition, but the chiral field generating
the multiplet has to be changed from~\eqref{eq:chiralfield} by conjugation
with a constant matrix
\begin{align}\label{}
    \phi=\frac{k}{4\pi}\tr(\begin{pmatrix}
                     0 & -1 \\
                     1 & 0 \\
                   \end{pmatrix}A^{-1}\begin{pmatrix}
                     0 & 1 \\
                     -1 & 0 \\
                   \end{pmatrix}
    B).
\end{align}
Then the corresponding F-term $\Phi=-\{G_{-1/2}^{-},[\bar G_{-1/2}^{-},\phi]\}$ gives
the boson-fermion interaction terms.

We can however go one step further. Looking at the free fermion
resolution~\eqref{eq:actionfreefermionic} we also have interaction terms
in the WZNW action for the bosonic subgroup. We can now go to the first
order formalism for the bosonic subgroup and write the bosonic Lagrangian
as first order kinetic terms plus bosonic screening charges. Our claim is
that the screening charges are also chiral perturbations. We will now show
this in the case of the new Manin decomposition.

The action for the bosonic subgroup~\eqref{eq:actionfreefermionic}
renormalized with the metric~\eqref{eq:glnngamma} takes the form:
\begin{align}\label{}
      S^{\textrm{WZNW}}_{ren}[g_B]&=- \frac{1}{4 \pi} \int_{\Sigma} d ^2z
 \langle g^{-1} \partial g , g^{-1} \bar \partial g \rangle_{\textrm{ren}}
 - \frac{1}{24 \pi} \int_{B} \langle g^{-1} d g ,
  [g^{-1} d g , g^{-1} d g ] \rangle_{\textrm{ren}}\nonumber\\
  &=\frac{1}{2\pi}\int_{\Sigma} d^2z \big[-(k-2)\del\phi^0_+\delbar\phi^0_++(k+2)\del\phi^0_-\delbar\phi^0_--2\del\phi^0_+\delbar\phi^0_--2\del\phi^0_-\delbar\phi^0_+ \nonumber\\
  &\phantom{=\frac{1}{2\pi}\int_{\Sigma} d^2z \big[}-(k-2)\del\phi_+\delbar\phi_+-(k-2)\del\bar\ga_+\delbar\ga_+e^{-2\phi_+}\nonumber\\
  &\phantom{=\frac{1}{2\pi}\int_{\Sigma} d^2z \big[}+(k+2)\del\phi_-\delbar\phi_-+(k+2)\del\bar\ga_-\delbar\ga_-e^{-2\phi_-}\big],
\end{align}
where we have chosen a parametrization of the Lie group valued field $g_B$ via a Gauss decomposition
\begin{align}\label{}
    g_B=e^{\ga_+ K_+^++\ga_-K_-^+}e^{\phi_+^0K_+^0+\phi_-^0K_-^0+\phi_+K_+^z+\phi_-K_-^z}e^{\bar\ga_+ K_+^-+\bar\ga_-K_-^-}.
\end{align}
We now introduce the auxiliary field $\be_\pm,\bar\be_\pm$. Remembering
corrections from the quantum measure -- which essentially cancels the
shift in levels for the two $\textrm{SU}(2)$ factors we get for the upper
$\textrm{SL}(2)$-part of the action
\begin{align}\label{}
    S_+=\frac{1}{2\pi}\int_{\Sigma} d^2z \left(-k\del\phi_+\delbar\phi_++\be_+\delbar\ga_++\bar\be\del\bar\ga_++\frac{1}{k-2}\be\bar\be e^{2\phi_+}+\frac{1}{4}\sqrt{h}\mathcal{R}^{(2)}\phi_+\right),
\end{align}
and similarly for the lower $\textrm{SL}(2)$ part. The bosonic currents in
this first order formalism are
\begin{align}\label{}
    J^{K_{+}^z}_B&=-2k\del\phi_++2\normord{\ga_+\be_+},\nonumber\\
    J^{K_{+}^+}_B&=\be_+,\nonumber\\
    J^{K_{+}^-}_B&=2k\del\phi_+\ga_+-(k-2)\del\ga_+-\normord{\be_+\ga_+\ga_+},
\end{align}
they can be inserted in the free fermion resolution~\eqref{eq:glnncurrents} to
give an expression for $G^-$ in~\eqref{eq:gminusnontypic}. For
completeness the anti-holomorphic currents are:
\begin{align}\label{}
    \bar J^{K_{+}^z}_B&=2k\delbar\phi_+-2\normord{\bar\ga_+\bar\be_+},\nonumber\\
    \bar J^{K_{+}^+}_B&=-2k\delbar\phi_+\bar\ga_++(k-2)\delbar\bar\ga_++\normord{\bar\be_+\bar\ga_+\bar\ga_+},\nonumber\\
    \bar J^{K_{+}^-}_B&=-\bar\be_+.
\end{align}
Using this, we compute that the screening charge $\Phi_B=\be\bar\be
e^{2\phi_+}$ is  the F-term of the supermultiplet obtained from the chiral
field
\begin{align}\label{}
    \phi_B=\frac{k^2}{k-2}c^{11}\bar c^{11}e^{2\phi_+},
\end{align}
and similarly for the screening charge of the lower $\trm{SL}(2)$ part.

In conclusion, we have shown that the whole action for the supergroup can
be written as simple first order kinetic terms (and background charges)
plus interaction terms in form of screening charges that can be written as
chiral F-terms. These are by construction exact in $G^-$ (and $\bar G^-$).
A practical consequence is that the cohomology of $G^-$ (and $\bar G^-$)
can be computed in free field theory.

\subsection{Comparison to string theory on $\adsst$}

In this section we show how our story is related to type IIB string theory
on $\adsst$.

String theory on $\adsst$ consists of bosons parameterizing the space
related to the group $\trm{SL}(2)\times\trm{SU}(2)\times \trm{U}(1)^4$,
their corresponding fermionic partners and the string ghosts. The total
central charge is zero.

Let us compare this to our approach for the case of the group
$\textrm{U}(1,1|2)_{-k}\times\trm{U}(1|1)\times\trm{U}(1|1)$. The bosonic
group that forms our starting point is then
$\trm{U}(1,1)\times\trm{U}(2)\times\trm{U}(1)^4$, and further we have the
corresponding decoupled fermionic partners. After the deformation with
$\gamma$, but before twisting the theory, this system also has central
charge zero. Remember that $\trm{U}(1,1)$ is equivalent to
$\trm{U}(1)\times\trm{SL}(2)$. Thus the bosonic group that we start with
has two $\trm{U}(1)$s plus their fermionic partners more than on the
string side. Bosonizing the fermions we thus have three extra scalars with
various background charges due to the deformation. On the string theory
side we, however, have the superconformal ghosts which consist of a
$\be\ga$- and a $bc$-ghost system. Bosonizing also gives three scalars
with background charges. So the field content is basically the same in
string theory and in our case of
$\textrm{U}(1,1|2)_{-k}\times\trm{U}(1|1)\times\trm{U}(1|1)$. We will now
show how to precisely embed string theory into our story.

\subsubsection*{$\adss$}
Let us first consider the string theory on $\adsst$ in detail
following~\cite{Giveon:1998ns}. The string theory is described by the
$\cN=1$ WZNW model on
$\trm{SL}(2)_{-k}\times\trm{SU}(2)_k\times\mathrm{U}(1)^4$ plus
superconformal ghosts. The negative level in $\trm{SL}(2)$ gives the
correct sign on the Cartan kinetic term and makes sure the total central
charge is zero. After decoupling the fermions the $\cN=1$ supercurrent
schematically takes the form
\begin{align}\label{eq:N1supercurrent}
    G^{\cN=1}_{\trm{string}}=J_a\chi^a-\frac{1}{6}{f_{ab}}^c\chi^a\chi^b\chi_c\, ,
\end{align}
where the index $a$ runs over the entire algebra, $J_a$ are the currents,
$\chi^a$ are the fermion partners from~\eqref{eq:fermionpartners}, and
${f_{ab}}^c$ are the structure constants of the entire algebra.

The string theory stress energy tensor splits in three parts corresponding
to the $\adss$, $\trm{T}^4$ and the ghosts
\begin{align}\label{}
    T_{\trm{string}}=T_{\trm{string}}^{\adss}+T_{\trm{string}}^{\trm{T}^4}+T_{\trm{string}}^{\trm{ghosts}},
\end{align}
and similarly for the rest of the superalgebra.

In appendix B of~\cite{Giveon:1998ns} the $\trm{U}(1)$ current,
$U_{\trm{string}}^{\adss}+U_{\trm{string}}^{\trm{T}^4}$, leading to the
chiral ring of string theory is written. Given $U_{\trm{string}}$ we can
find $G^\pm_{\trm{string}}$ using that they have opposite $\trm{U}(1)$
charge and
$G^{\cN=1}_{\trm{string}}=G^+_{\trm{string}}+G^-_{\trm{string}}$. Here and
in the following we write the basis of $\trm{SU}(2)$ in the form of
$\trm{SL}(2)$, i.e. we use generators
\begin{equation}\label{}
t^\pm=t^x\pm i t^y.
\end{equation}
We will use our notation for the basis of $\trm{SL}(2)\times\trm{SU}(2)$
as in~\eqref{eq:newbasisgl22} i.e. $K^{\pm,z}_+$ spans $\trm{SL}(2)$ and
$K_-^{\pm,z}$ span $\trm{SU}(2)$.

The $\adss$ part of the $\trm{U}(1)$ from~\cite{Giveon:1998ns} then reads
\begin{equation}\label{eq:Ustring}
    U_{\trm{string}}^{\adss}=\frac{1}{k}J^{K_+^z}-\frac{1}{k}J^{K_-^z}+\frac{1}{2k}\chi^{K_+^z}\chi^{K_-^z}-\frac{k+2}{k^2}\chi^{K_+^+}\chi^{K_+^-}+\frac{k-2}{k^2}\chi^{K_-^+}\chi^{K_-^-}\, .
\end{equation}
We can now ask what the expressions for $G^\pm_{\trm{string,\,}\adss}$
are. To find these we simply note that the $\trm{U}(1)$ current
corresponds to the Manin decomposition
\begin{align}\label{eq:stringmanin}
    \ap&=\{K_+^z-K_-^z,\,K_-^-,\,-K_+^-\} \nonumber\\
    \am&=\{-\frac{1}{4 k}(K_+^z+K_-^z),\,\frac{1}{k}K_-^+,\,\frac{1}{k}K_+^+\}.
\end{align}
The currents $G^\pm_{\trm{string,\,}\adss}$ then follow directly
from~\eqref{eq:gplusminus}. Note that this Manin decomposition is very
symmetric, and the $G^+$- and $G^-$-chiral ring are dual (conjugately
related). Further, $\am$ is exactly the same as we have in the
$\textrm{U}(1,1|2)_{-k}$ case with the overall $U(1)$s removed for the
alternative Manin decomposition in~\eqref{eq:gl22manin2}. So if we
identify our fermions $\chi_i,\chi^i$ with the corresponding string
fermions we get the same $G^-$ current, and actually also the same $U$
current
\begin{align}\label{}
    G^-_{\trm{string,\,}\adss}&=G^-|_{\trm{SL}(2)\times\trm{SU}(2)},\\
        U_{\trm{string,\,}\adss}&=U|_{\trm{SL}(2)\times\trm{SU}(2)},
\end{align}
however the two $G^+$ currents differ since the Manin bases  $\ap$ are
different.\footnote{Since $\ap$ is simpler in string theory than in our
case, the supersymmetry algebra have a two-dimensional space of
deformations instead of the one-dimensional in our case.} In the basis of
$K^{\pm,z}_\pm$ the identification of the fermions are trivial except for
\begin{align*}
    \chi^{K_-^-}_{\trm{string}}=\chi^{K_-^-}-\chi^{K_+^+},\qquad \chi^{K_+^-}_{\trm{string}}=\chi^{K_+^-}-\chi^{K_+^-},
\end{align*}
due to the difference in the $\ap$ bases. This change in fermions
corresponds to an isometry and hence do not change the stress-energy
tensor. However, it is not a Lie algebra automorphism and thus cannot be
extended to the currents which we identify trivially. Also, this means
that $G^{\cN=1}_{\trm{string}}$ does not get mapped into
$G^{\cN=1}=G^++G^-$. Hence $G^-$ and $U$ can get mapped into each other
while $G^+$ differs. The difference is
\begin{align}\label{}
    G^+_{\trm{string,\,}\adss}&=G^+|_{\trm{SL}(2)\times\trm{SU}(2)}+\frac{1}{k}J^{K_+^+}\chi^{K_-^+}-\frac{1}{k}J^{K_-^+}\chi^{K_+^+}-\frac{1}{k^2}(\chi^{K_+^z}-\chi^{K_-^z})\chi^{K_+^+}\chi^{K_-^+}\, .
\end{align}
In other words, we have constructed another $\cN=2$ supersymmetry of the
string theory that only differ in the $G^+$ current. This difference is
important to achieve that $G^+$ after deformation and twist is a current
of the $\textrm{U}(1,1|2)_{-k}$ model.

Further, we have to do the $\ga$ deformation~\eqref{eq:aldeformmanin2}.
For the $\ads_3$ part this is just like the type of spectral flows
suggested in~\cite{Maldacena:2000hw}. This is then extended to also
include the $\trm{S}^3$ sector and to deform the fermions to preserve the
$\cN=2$ supersymmetry. This part of the deformation does not change the
central charge. Let us again stress that the deformation only changes the
weights in the chiral ring.

\subsubsection*{Ghosts and $\trm{T}^4$}

We want to compare the ghosts and $\trm{T}^4$ of string theory with the
$\trm{U}(1)$s in the
$\textrm{U}(1,1|2)_{-k}\times\trm{U}(1|1)\times\trm{U}(1|1)$. To study
this we need to add the generators of the bosonic bases of $\trm{U}(1|1)$
to the Manin decomposition~\eqref{eq:gl22manin2}. This means that the
$\cN=2$ supersymmetric theory of the $\trm{U}(1)$s is described by $J_i$,
$J^i$ and the superpartners $\chi_i$, $\chi^i$ where $i=1,5,6$. The
supersymmetric algebra after the deformation given
by~\eqref{eq:aldeformmanin2} is then
\begin{align}\label{eq:ourN2u1s}
\begin{split}
    G^+_\ga|_{\trm{U(1)s}}&=J_i\chi^i+\del(2\chi^1-\chi^5-\chi^6), \\
    G^-_\ga|_{\trm{U(1)s}}&=J^i\chi_i+\del(\chi_1-\half\chi_5-\half\chi_6), \\
    U_\ga|_{\trm{U(1)s}}&=\chi^i\chi_i+J_1-2J^1-\half J_5+J^5-\half J_6+J^6, \\
    T_\ga|_{\trm{U(1)s}}&=J_iJ^i+\half(\del\chi^i\chi_i-\chi^i\del \chi_i)+\del(\half J_1+J^1-\fourth J_5-\half J^5-\fourth J_6-\half J^6).
\end{split}
\end{align}
Here we have performed a rescaling of the fermions and currents to absorb
the levels, but keeping the kinetic terms. The levels can in principle be
different for the $\trm{U}(1|1)$s.

We now turn to the $\trm{T}^4$ part of string theory. To anticipate the
embedding into our theory we denote the currents by $\hat J_i,\hat J^i$
and the fermions by $\hat \chi_i,\hat \chi^i$ where $i=5,6$. The $\cN=2$
algebra is given standardly by.\footnote{Note that~\cite{Giveon:1998ns}
has an extra overall $i$ in the $U$ current.}
\begin{align}\label{eq:stringN2T4}
\begin{split}
    G^+_{\trm{string,\,}\trm{T}^4}&=\hat J_i\hat \chi^i, \\
    G^-_{\trm{string,\,}\trm{T}^4}&=\hat J^i\hat \chi_i, \\
    U_{\trm{string,\,}\trm{T}^4}&=\hat \chi^i\hat \chi_i, \\
    T_{\trm{string,\,}\trm{T}^4}&=\hat J_i\hat J^i+\half(\del\hat \chi^i\hat \chi_i-\hat \chi^i\del\hat  \chi_i).
\end{split}
\end{align}

Finally, we consider the ghost system which consists of a $\be\ga$-system
of central charge $11$ and a $bc$-system of central charge $-26$. The
$\cN=1$ supercurrent of the ghost system is given in
e.g.~\cite{Polchinski:1998rr}.\footnote{We correct the formula with an
$i$.} Without bosonization the only possible extension to an $\cN=2$
algebra (without bosonization and up to swapping of $G^\pm$) takes the
following form
\begin{align}
\begin{split}
    G^+_{\trm{string,\,}\trm{ghost}}&=-2ib\ga, \\
    G^-_{\trm{string,\,}\trm{ghost}}&=-i(\del\be)c-i\frac{3}{2}\be\del c, \\
    U_{\trm{string,\,}\trm{ghost}}&=-2\normord{bc}-3\normord{\ga\be}, \\
    T_{\trm{string,\,}\trm{ghost}}&=-\normord{\del b c}-2\normord{b\del c}-\frac{1}{2}\normord{\be\ga}-\frac{3}{2}\normord{\be\del\ga}.
\end{split}
\end{align}
We now bosonize the ghosts into three scalars, the $bc$-system in the
ordinary way $b=e^{-\rho_1}$, $c=e^{\rho_1}$ and the $\be\ga$-system we
bosonize oppositely as normal $\ga=ie^{-\rho_2+\rho_3}\del\rho_3$,
$\be=ie^{\rho_2-\rho_3}$. The bosonized currents are
\begin{align}
\begin{split}
    G^+_{\trm{string,\,}\trm{ghost}}&=2e^{-\rho_1-\rho_2+\rho_3}\del\rho_3, \\
    G^-_{\trm{string,\,}\trm{ghost}}&=e^{\rho_1+\rho_2-\rho_3}\del(\frac{3}{2}\rho_1+\rho_2-\rho_3) \\
    U_{\trm{string,\,}\trm{ghost}}&=\del(2\rho_1+3\rho_2), \\
    T_{\trm{string,\,}\trm{ghost}}&=\frac{1}{2}\del\rho_1\del\rho_1+\frac{3}{2}\del^2\rho_1-\frac{1}{2}\del\rho_2\del\rho_2+\frac{1}{2}\del\rho_3\del\rho_3+\del^2\rho_2+\frac{1}{2}\del^2\rho_3.
\end{split}
\end{align}
To be able to compare we re-fermionize this system. We define $\hat
\chi_1=\frac{3}{2}e^{\rho_1+\rho_2-\rho_3},\hat
\chi^1=\frac{2}{3}e^{-\rho_1-\rho_2+\rho_3}$, $\hat J_1=3(\rho_1+\rho_2)$
and $\hat J^1=\frac{1}{3}(-\rho_2+\rho_3)$. We then arrive at
\begin{align}\label{eq:stringN2ghosts}
\begin{split}
    G^+_{\trm{string,\,}\trm{ghost}}&=\hat J_1\hat \chi^1+3\del\hat\chi^1, \\
    G^-_{\trm{string,\,}\trm{ghost}}&=\hat J^1\hat \chi_1+\del\hat\chi_1, \\
    U_{\trm{string,\,}\trm{ghost}}&=\hat \chi^1\hat \chi_1+\del\hat J_1-3\del\hat J^1, \\
    T_{\trm{string,\,}\trm{ghost}}&=\hat J_1\hat J^1+\half(\del\hat \chi^1\hat \chi_1-\hat \chi^1\del\hat  \chi_1)+\half\del\hat J_1+\tfrac{3}{2}\del\hat J^1.
\end{split}
\end{align}
Note that this could also easily be put into a form that is symmetric in
$\G^\pm$ again showing that the $G^\pm$-chiral rings of string theory are
in one-to-one correspondence.

We can now get a direct match of our superalgebra~\eqref{eq:ourN2u1s} for
the $\trm{U}(1)$s and the string superalgebra for the ghosts and
$\trm{T}^4$ in equations~\eqref{eq:stringN2T4}
and~\eqref{eq:stringN2ghosts} using the dictionary
\begin{align*}
    \hat\chi_{(i)}&=A\chi_{(i)},& \hat\chi^{(i)}&=(A^{-1})^T\chi^{(i)}\, ,\\
    \hat J_{(i)}&=A J_{(i)},& \hat J^{(i)}&=(A^{-1})^T J^{(i)}\, ,
\end{align*}
where $\chi_{(i)}^T$ is the vector $(\chi_1,\chi_5,\chi_6)$ etc. and $A$
is the matrix
\begin{align*}
    A=\begin{pmatrix}
        1 & -\half & -\half \\
        \half & 1 & 0 \\
        \half & 0 & 1 \\
      \end{pmatrix}\, .
\end{align*}

To sum up, we have shown in this section that the superalgebra we have
after deformation, i.e. right before twisting, in the case of
$\textrm{U}(1,1|2)_{-k}\times\trm{U}(1|1)\times\trm{U}(1|1)$ only differs
from the standard string theory algebra in the $G^+$ part, and that our
algebra is simply another choice of $\cN=2$ algebra.

\subsection{Boundary actions and the Warner problem}

In this section, we want to understand boundary actions in supergroup WZNW models
using methods of world-sheet supersymmetric theories.
Warner explained how to find B-type boundary actions in superconformal field thoeries \cite{Warner:1995ay}, we use \cite{Brunner:2003dc, Hori:2004zd} as references.
Boundary conformal field theory on supergroups hasbeen investigated in
\cite{Creutzig:2007jy,Quella:2007sg,Creutzig:2008ag,Creutzig:2008ek,Creutzig:2006wk,Creutzig:2008an,Creutzig:2009zz}.

Lie supergroup bulk WZNW models can be well treated in the free fermion formalism.
A similar method is desired for the boundary theories. The problem is to find
 the appropriate boundary action. So far only in GL(1$|$1) \cite{Creutzig:2008ek} and in OSP(1$|$2)
\cite{Creutzig:2010zp} this problem could be solved.
On the other hand, knowing the boundary action was essential in solving these models.

Boundary conformal field theory in WZNW models is characterized by boundary conditions that preserve
the current algebra in addition to conformal symmetry. This is the case if and only if the holomorphic
and anti-holomorphic currents are glued together at the boundary with a metric preserving
automorphism $\omega$ of the underlying horizontal subalgebra
\begin{equation}
 J(z) \ = \ \omega\bigl(\bar J(\bar z)\bigr) \qquad\text{for} \ z\ = \ \bar z\, .
\end{equation}
A consequence of these conditions is that the group valued field that describes the sigma model maps the
boundary of the world-sheet to a twisted (super) conjugacy class $C_a^\omega$,
\begin{equation}
   C_a^\omega \ = \ \{\, g = ha\omega(h^{-1}) \,|\, h\,\in\, G\,\}\,.
\end{equation}
The constant element $a$ is parameterizing the position of the brane.

Now, we saw that the bulk GL(N$|$N) WZNW model can be obtained by twisting
an $\mathcal{N}=(2,2)$ superconformal model. Moreover, the boson-fermion
interaction term is an F-term. In this section, we want to consider the
boundary theories with gluing automorphism $\omega$ being minus one times
the transpose in our matrix representation. This implies that our matrix
valued fields $A$ and $B$ in GL(N) have the form
\begin{equation}\label{eq:conjugacy}
 A \ = \ CA_0C^t \qquad\text{and}\qquad B \ = \ DB_0D^t
\end{equation}
for some GL(N) valued fields $C$ and $D$ and constants matrices $A_0$ and $B_0$.

We want to find a boundary action that preserves the superconformal
symmetry. This problem is often referred to as the Warner problem. Its
solution requires additional fermionic boundary degrees of freedom as well
as a factorization of the super potential into boundary super potentials.
Let us review this situation for B-branes in Landau-Ginzburg models. The
world-sheet of a Landau-Ginzburg model has two complex fermionic
coordinates. The action of the model is given by a D-term and an F-term,
$\mathcal{L}=\mathcal{L}_D+\mathcal{L}_F$, which are both by construction
invariant under supersymmetry transformations. If one integrates the
fermionic coordinates, the F-term looks as follows
\begin{equation}
\begin{split}
\mathcal{L}_F\ = \ \frac{1}{4}|W'|^2 + \frac{1}{2}W''\psi_+\psi_--\frac{1}{2}\bar W''\bar\psi_+\bar\psi_-\, .
\end{split}
\end{equation}
A simple choice of D-term is, after integrating the fermionic coordinates,
the Lagrangian of free bosons and fermions
\begin{equation}
\begin{split}
\mathcal{L}_D\ = \ \del\bar\phi\bar\del\phi+\del\phi\bar\del\bar\phi+
 \bar\psi_-\del\psi_-+\psi_-\del\bar\psi_-+\bar\psi_+\bar\del\psi_++\psi_+\bar\del\bar\psi_+\, .
\end{split}
\end{equation}
There exist two families of boundary conditions that preserve half of the
supersymmetry, which are called A- and B-boundary conditions. We are
interested in the second case. First, if one sets the super potential $W$
to zero, B-type supersymmetry is preserved by introducing the following
boundary term
\begin{equation}
 S_{0,\text{bdy}} \ = \ \frac{i}{4}\int d\tau\ \bar\theta\eta-\bar\eta\theta\, ,
\end{equation}
where $\eta=\psi_-+\psi_+$ and $\theta=\psi_--\psi_+$. In this free theory
this amounts to the boundary conditions $\psi_-=\psi_+$ and
$\bar\psi_-=\bar\psi_+$ for the fermions. For non-zero super potential
$W$, the supersymmetry variation of the action gives a boundary term of
the form
\begin{equation}
 \delta_{\text{susy}}(S+S_{0,\text{bdy}}) \ = \ \frac{i}{2}\int d\tau \ \epsilon\bar\eta\bar W´ +\bar\epsilon\eta W´\, .
\end{equation}
Finding a boundary term whose variation cancels this contribution is the
Warner problem. Its solution is given by introducing a fermionic boundary
super field. After integrating the odd world-sheet coordinates the
boundary term is
\begin{equation}\label{bdyLG}
 S_{\text{bdy}}\ = \ \int d\tau\ i\bar\pi\del_\tau\pi-\frac{1}{2}\bar JJ
-\frac{1}{2}\bar EE+\frac{i}{2}\pi\eta J^\prime
+\frac{i}{2}\bar\pi\bar\eta \bar J^\prime
-\frac{1}{2}\bar\pi\eta E^\prime
-\frac{1}{2}\pi\bar\eta \bar E^\prime\, .
\end{equation}
Here $J(\phi)$ and $E(\phi)$ are boundary potentials, prime denotes
derivative with respect to $\phi$, and $\pi$ is the new boundary fermion.
Its variation under supersymmetry is
\begin{equation}\label{eq:susyfermion}
 \delta_{\text{susy}}\pi \ = \ -i\epsilon\bar J - \bar\epsilon E \qquad, \qquad
 \delta_{\text{susy}}\bar\pi \ = \ i\bar\epsilon J - \epsilon \bar E\, .
\end{equation}
The total action is now invariant under supersymmetry variation if and only if
\begin{equation}
 W \ = \ EJ+\text{constant}\, .
\end{equation}

In the case of the $\trm{U}(1)\times\trm{SU}(2)$ WZNW model a superfield
formulation of the theory is known~\cite{Ivanov:1994ec}. This motivates us
to use an analogous method to solve the Warner problem in our case. We
already saw in the last section that the bulk superpotential is
\begin{equation}
 W(A,B)\ = \ \tr(A^{-1}B)\qquad, \qquad\bar W\ = \ 0\, .
\end{equation}
Here $A$ and $B$ are the two GL(N) (matrix) valued fields describing the GL(N)$\times$GL(N) WZNW model.
Recall that we choose gluing conditions for the currents that force the matrix valued fields to have the
form $A=CA_0C^t$ and $B=DB_0D^t$ at the boundary, where $C$ and $D$ are some matrix valued fields.

Further, since $\bar W=0$, we should have $\bar J=\bar E=0$. The field
$\eta$ is in our case $\eta=b=\bar b^t$ and is a gl(N) matrix valued
fermionic field. Since we identify $b$ with $\bar b^t$, we also want to
identify $\bar \pi$ with $\pi$ in some way. Since $\bar\pi \pi$ should be
a scalar field we take
\begin{equation}
 i\bar\pi \ = \ \tr(B_0\pi^tA_0^{-1}\ \cdot\ )
\end{equation}
for two constant GL(N) matrices $A_0$ and $B_0$. Then the supersymmetry variation \eqref{eq:susyfermion} forces to also identify
\begin{equation}
 J \ = \ \tr(B_0E^tA_0^{-1}\ \cdot\ )\, .
\end{equation}
We define $E=C^{-1}D$ for two GL(N) valued boundary fields $C$ and $D$,
such that the invariant vector fields act as
\begin{equation}
\del_{\sigma\delta}E \ = \ C^{-1}E^{\sigma\delta}D\, .
\end{equation}
Moreover, we have $JE=W$ from~\eqref{eq:conjugacy} as required.
Then the boundary action \eqref{bdyLG} takes the form
\begin{equation}\label{bdyLG2}
 S_{\text{bdy}}\ = \ \int d\tau\ \tr(B_0\pi^tA_0^{-1}\del_\tau\pi)-\frac{i}{2}\tr(B_0D^tb^t(C^{-1})^tA_0^{-1}\pi)
+\frac{i}{2}\tr(B_0\pi^tA_0^{-1}C^{-1}bD)\, .
\end{equation}
We believe that this is the correct action of the boundary model, and it
is indeed in the example of GL(1$|$1) \cite{Creutzig:2008ek}. In order to
prove this statement rigourously one has to show that the boundary
potential is a screening charge for the currents. We have not proven this
in general, but it is straightforward to address this issue in the example
one wants to study.

\section{Summary and Outlook}

We have shown that the B-twist of the world-sheet supersymmetric
$\trm{GL}(N)\times\trm{GL}(N)$ WZNW model perturbed by a truly marginal
operator, an F-term, is embedded in the $\trm{GL}(N|N)$ WZNW model.
Moreover, the supercurrents, as well as the U$(1)$-current, are
expressions in terms of Lie superalgebra currents. Further, we have seen
that the principal chiral field is a D-term.

We then applied these results. In the example of the GL$(2|2)$ WZNW model
we have shown that the action consists of a free kinetic term plus two
F-terms. One term couples bosons to fermions while the other gives
screening charges for the bosonic subgroup $\trm{GL}(2) \times
\trm{GL}(2)$. The important consequence is that one can compute the
cohomology of the current $G^-$ in free field theory.

Another application is the relation to type IIB string theory on $\adsst$.
Here we have shown that string theory is related by our procedure to the
$\textrm{U}(1,1|2)_{-k}\times\trm{U}(1|1)\times\trm{U}(1|1)$ WZNW model,
and that the current $G^-$ is indeed the standard choice of supersymmetry
charge. The current $G^+$ is a novel choice which is important for our
relation.

We suspected a relation between the GL$(N|N)$ WZNW models and $\cN=(2,2)$
world-sheet supersymmetric theories because the GL(1$|$1) boundary WZNW
model action is very similar to what one finds in boundary superconformal
field theories. Knowing the boundary action is an essential aide in
solving the model, and hence it was important that we could apply the
techniques of superconformal field theory to find boundary actions for our
models.

Two tasks remain to be done. Firstly, to use the protected sectors we have
found and compute the corresponding cohomology and their correlation
functions. It would be particularly interesting to do this for world-sheets of higher
genus. Secondly, to verify the boundary actions in an example and use them
to compute correlation functions in the boundary theory.

There are also possible generalizations. A topological twist of an
$\cN=(2,2)$ world-sheet supersymmetric theory has central charge zero, and
thus can only correspond to a supergroup WZNW model whose superdimension
is zero. We already checked that the B-twist of the superconformal SL(2)
$\times$ U(1) WZNW model is related to the supergroup SL(2$|$1) WZNW
model, but have not yet discovered other examples. Also note that there
exist world-sheet superconformal supergroup WZNW models
\cite{Creutzig:2009fh}. One might expect that their B-twists in some cases
can be related to WZNW models of orthosymplectic supergroups, and this
should be investigated.

\subsection*{Acknowledgements}

We would like to thank Manfred Herbst, Ingo Kirsch, Hubert Saleur, David Ridout and especially Volker
Schomerus for helpful discussions.


\providecommand{\href}[2]{#2}\begingroup\raggedright\endgroup

\end{document}